\documentclass[prx,twocolumn,showpacs,preprintnumbers,amsmath,amssymb]{revtex4-1}
\usepackage{graphicx}
\usepackage{ulem}

\begin{document}
\title{  Quantum incommensurate Skyrmion  crystals and Commensurate to In-commensurate transitions in cold atoms and 
materials with  spin orbit couplings in a Zeeman field }
\author{ Fadi Sun$^{1,2,3}$, Jinwu Ye,$^{1,3,4}$ and Wu-Ming Liu$^{2}$  }
\affiliation{
$^{1}$Department of Physics and Astronomy, Mississippi State University, MS, 39762, USA \\
$^{2}$Beijing National Laboratory for Condensed Matter Physics,
Institute of Physics, Chinese Academy of Sciences,
Beijing 100190, China   \\
$^{3}$Key Laboratory of Terahertz Optoelectronics, Ministry of Education, Department of Physics, Capital Normal University, Beijing, 100048, China \\
$^{4}$  Kavli Institute of Theoretical Physics, University of California, Santa Barbara, Santa Barbara, CA 93106   }

\date{\today }

\begin{abstract}
 In this work, we study strongly interacting spinor atoms in a lattice subject to a 2 dimensional (2d) anisotropic Rashba
 type of spin orbital coupling (SOC) and an Zeeman field. We find the interplay between the Zeeman field and the SOC
 provides a new platform to host rich and novel classes of quantum commensurate and in-commensurate phases, excitations and phase transitions.
 These commensurate phases include two collinear states at low and high Zeeman field,
 two co-planar canted states at Mirror reflected SOC parameters respectively.
 Most importantly, there are non-coplanar incommensurate Skyrmion (IC-SkX) crystal phases surrounded by the 4 commensurate phases.
 New excitation spectra above all the 5 phases, especially on the IC-SKX phase are computed.
 Three different classes of quantum commensurate to in-commensurate transitions from the IC-SKX to its 4 neighboring commensurate phases are identified.
 Finite temperature behaviors and transitions are discussed.
 The critical temperatures of all the phases can be raised above that reachable
 by current cold atom cooling techniques simply by tuning the number of atoms $ N $ per site.
 In view of recent impressive experimental advances in generating 2d SOC for cold atoms in optical lattices, these new many-body phenomena
 can be explored in the current and near future cold atom experiments.
 Applications to various materials such as MnSi, Fe$_{0.5}$Co$_{0.5}$Si,
 especially the complex incommensurate magnetic ordering in Li$_2$IrO$_3$ are given.
\end{abstract}

\maketitle

\section{ Introduction }
Quantum phases and phase transitions in quantum spin systems
have been an important and  vigorous research field in material science for many decades
\cite{sachdev,aue,scaling}. However, so far, most of these quantum phases are  collinear phases
in a bipartite lattice or co-planar commensurate
phases in a geometrically frustrated lattice. The associated quantum phase transitions are commensurate to commensurate (C-C) ones.
During the last decade, the investigation and control of spin-orbital coupling (SOC) \cite{ahe} have become  the subjects of
intensive research in both condensed matter and cold atom systems
after the discovery of the topological insulators
\cite{kane,zhang}.
In the condensed matter side, there are increasing number of
new quantum materials with significant SOC,
including several new 4d or 5d transition metal oxides
and heterostructures of transition metal systems
\cite{kitpconf}.
In the cold atom side, there were series advances to generate Abelian gauge flux and quantum spin Hall effects in optical lattices
\cite{stagg1,stagg2,stagg3,stagg4,uniform1,uniform2,uniform3,newexp,newexp2,sorev}.
Several groups worldwide \cite{soexp,sofermigas,sobecgas} have also successfully generated
a 1D  (SOC) to neutral atoms. However, one of the main limitations to extend 1D SOC to a 2D SOC is the associated heating rates.
Recently, there are also some advances \cite{expk40,expk40zeeman,clock,2dsocbec} to overcome this difficulty
in generating 2D Rashba SOC for cold atoms in both continuum and optical lattices and also in a Zeeman field.
Most recently, a long-lived SOC gas of the high magnetic fermionic element dysprosium to eliminate the heating due to the spontaneous emission,
has been created in \cite{ben}. In view of these recent experimental advances,
novel superfluid or magnetic phenomena due to the interplay among
tunable interactions, SOC and a Zeeman field are ready to be investigated in near future experiments on both fermion and spinor BEC.
It becomes topical and important to investigate what would be new phenomena due to such an interplay in both cold atoms and
condensed matter systems.



In a recent work \cite{rh}, we studied interacting spinor bosons at integer fillings loaded in a square optical lattice in the presence of non-Abelian gauge fields. In the strong coupling limit, it leads to the  the spin $ S $ Rotated Ferromagnetic Heisenberg model
which is a new class of quantum spin models to describe quantum magnetisms in cold atom systems or some materials with strong SOC \cite{wu,classdm1,classdm2}.
Along a anisotropic line of 2d SOC, we identified a new spin-orbital entangled commensurate ground state:
the Z-x state.
It supports not only commensurate magnons (C-C$_0$,C-C$_{\pi}$),
but also a new gapped elementary excitation: in-commensurate magnon (C-IC).
The C-IC magnons may become the seeds to drive possible new classes of quantum C-IC transitions under various external probes.
In this paper, we study possible dramatic effects of
an external Zeeman field $H$ applied to the Rotated Ferromagnetic Heisenberg model.
We find that the interplay among the strong interactions, SOC and the Zeeman field leads to a
whole new classes of magnetic phenomena in quantum phases ( especially the non-coplanar incommensurate Skyrmion  crystals (IC-SkX)  ),
excitation spectra ( especially inside the IC-SkX ), quantum  phase transitions ( especially the quantum Commensurate to incommensurate (C-IC) transitions ),
which have wide and important applications in both cold atoms and various materials with SOC.
Our main results are summarized in  Fig.1 and Fig.2.
We also discuss the finite temperature behaviors and finite temperature phase transitions above the $ T=0 $ quantum phases in Fig.1 and Fig.2.
Particularly, we point out that any spin $ S=N/2 $ of the RFH can be simply achieved by tuning the number of atoms $ N $ per site,
the critical temperatures of all the phases  $ T_c/J \sim 2S=N $ at 2 dimension can be easily increased above that reachable
by current cold atom cooling techniques.
In view of recent impressive experimental advances in generating 2d SOC for cold atoms in optical lattices, these new many-body phenomena
can be explored in the current and near future cold atom experiments. The SOC materials with a total spin $ J=1/2 $
automatically fall into the strong coupling regime.
We also discuss the applications of Fig.1 and Fig.2 to various materials such as MnSi, Fe$_{0.5}$Co$_{0.5}$Si,
especially the complex incommensurate magnetic ordering in Li$_2$IrO$_3$.



In the previous work \cite{rh}, the authors studied interacting
spinor bosons at integer fillings $ N $ hopping in a square optical
lattice subject to any linear combinations of Rashba and
Dresselhaus spin-orbit coupling (SOC). In the strong-coupling
limit, it leads to the spin $ S=N/2 $  Rotated ferromagnetic Heisenberg model
(RFHM)  at a zero Zeeman field Eq.\ref{rhgeneral}, which is
a new class of quantum spin models to describe quantum
magnetisms in cold-atom systems or some materials with
strong SOC. The spin $ S=N/2 $ Rotated Ferromagnetic Heisenberg model \cite{rh}
at a generic SOC parameters $ ( \alpha, \beta) $ in a Zeeman field $ \vec{H} $ is:
\begin{eqnarray}
	\mathcal{H}_{RH} & = & -J\sum_i
	[\mathbf{S}_i R(\hat{x},2\alpha)\mathbf{S}_{i+\hat{x}}
	+\mathbf{S}_i R(\hat{y},2\beta)\mathbf{S}_{i+\hat{y}}]   \nonumber  \\
    & - & \vec{H} \cdot \sum_i \vec{S}
\label{rhgeneral}
\end{eqnarray}
where  $H$ is the Raman laser induced Zeeman field
\cite{stagg1,stagg2,stagg3,stagg4,uniform1,uniform2,uniform3,newexp,newexp2}.

Here, we also follow \cite{rh} to take a "divide and conquer" strategy.
We first explore new and rich quantum phenomena along the
solvable line $\alpha=\pi/2, 0<\beta<\pi/2$.
Then starting from the deep knowledge along the solvable line,
we will try to investigate the quantum phenomena at generic $(\alpha,\beta)$.
As shown in \cite{rh},  the RFHM along the line at $ H=0 $ has the translational symmetry,
the time reversal $ {\cal T} $, the three spin-orbital coupled $ Z_2 $
symmetries $ {\cal P}_x, {\cal P}_y, {\cal P}_z $. Most importantly, it also owns
a hidden spin-orbital coupled $ U(1)_{soc} $ symmetry generated by
$ U_1(\phi)=e^{ i \phi \sum_{i} (-1)^x S^{y}_i } $.
As shown in \cite{rh}, the RFHM along the solvable line
has an exact ground state $ Y-x $ state with 2-fold degeneracy. So
in this paper, we focus on studying the phenomena in the Zeeman field along the longitudinal $ y $ direction.
If one adds a staggered Zeeman field coupled to the conserved quantity
$ \sum_{i} (-1)^x \sum_i S_i^y $, then it will pick one of the two degenerate $ Y-x $ state at $ h=0 $.
There is no phase transitions in this case.
The $ H $ breaks the $ {\cal T}, {\cal P}_x, {\cal P}_z $ symmetries,
but still keeps the translation, $ {\cal P}_y $,
the combinations $ {\cal T}{\cal P}_x, {\cal T}{\cal P}_z $ and the hidden $ U(1)_{soc} $ symmetry.
It can be shown that under the Mirror transformation $ {\cal M} $ which consists of the local rotation
$\tilde{\mathbf{S}}_{i} =R(\hat{x},\pi ) R(\hat{y},\pi n_2) \mathbf{S}_{i}$
followed by a Time reversal transformation $ {\cal T } $,
$ (\beta, h ) \rightarrow ( \pi/2 - \beta, h ) $.

Obviously, due to the lacking of the spin $ SU(2) $ symmetry in Eqn.\ref{rhgeneral},
applying the Zeeman field along the two transverse directions  $ H_x $ and $ H_z $ lead to quite different phenomena
and will be presented in separate publications \cite{rhtran}.
Rotated Anti-ferromagnetic model(RAFM) will show quite different behaviors \cite{rafm,rafm2}.

\begin{figure}
\includegraphics[width=7.5cm]{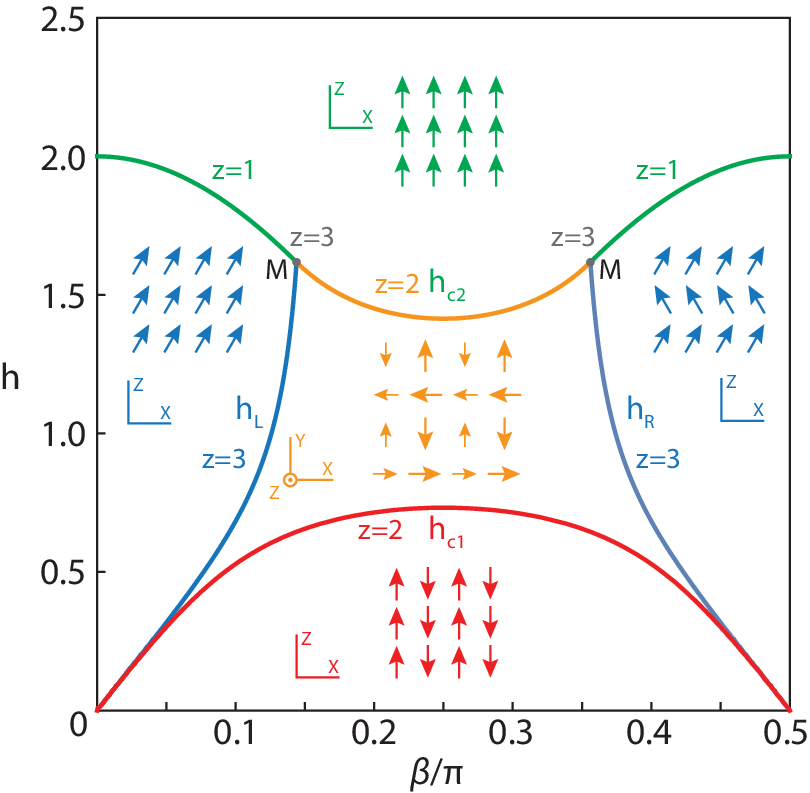}
\caption{ Quantum Phases and phase transitions of RFHM in a Zeeman field Eq.\eqref{rhh}.
 Below $ h_{c1} $ is the spin-orbital correlated (collinear) Z-x state.
 Above $ h_{c2} $ is the (collinear) Z-FM state. Note the three different pieces of $ h_{c2} $.
 On the left, $ h_L $ is one canted (co-planar) state. On the right, $ h_R $ is another
 canted (co-planar) state. Surrounded by the four commensurate phases is the in-commensurate Skyrmion crystal (non-coplanar) phase (IC-SkX)
 with non-vanishing Skyrmion density. At $\beta=\pi/4$, the IC-SkX reduces to a $2\times4$ commensurate SkX where
 only the spins (with two different lengths) in the $XY$ plane are shown, its $Z$ component is shown in Fig.\ref{pi4}a.
 There is a multi-critical ( M )  point where the ( collinear ) Z-FM, the ( co-planar ) canted phase and the ( non-co-planar ) IC-SkX phase meet.
 The phases on the left $ \beta < \pi/4 $ are related to the right $ \beta > \pi/4 $ by the Mirror transformation. The center $ \beta = \pi/4 $ respects the Mirror symmetry.  The excitation spectra above all these quantum phases, especially those inside the IC-SkX are worked out.
 The quantum phase transitions between (among) these phases with the dynamic exponents $z=1$, $z=2$ and the anisotropic one $(z_x=1,z_y=3)$
 ( abbreviated by $z=3$ in the figure ), especially the three different classes of quantum commensurate to In-commensurate transitions
 are discussed in the text. }
\label{globalphase}
\end{figure}

After rotating spin $Y$  axis to $Z$ axis by the global rotation  $R_x(\pi/2)$
(or equivalently, one can just put $ \beta \sigma_z $  along the $ y $ bonds in the square lattice ),
the Hamiltonian Eqn.\ref{rhgeneral} along the line $ ( \alpha=\pi/2, 0<\beta<\pi/2) $
in the $ H $ field along $ y $ direction can be written as:
\begin{eqnarray}
	\mathcal{H} & = & -J\sum_i[\frac{1}{2}(S_i^+S_{i+x}^+ + S_i^-S_{i+x}^-)
	-S_i^zS_{i+x}^z    \nonumber  \\
	 & + & \frac{1}{2}(e^{i2\beta}S_i^+S_{i+y}^-+e^{-i2\beta}S_i^-S_{i+y}^+)
	+S_i^zS_{i+y}^z]   \nonumber  \\
	& - & H \sum_i S_i^z
\label{rhh}
\end{eqnarray}
where the Zeeman field $H$ is along the $\hat{z}$ direction after the global rotation.
In the following, we take $ 2SJ $ as the energy unit, so all the all the physical quantities
such as the Zeeman field $ H $, the magnon dispersion $ \omega_k $ and the gap $ \Delta $  will
be dimensionless after taking their ratios over $ 2SJ $.
We will first focus on the left half of Fig.\ref{globalphase} with $ 0 < \beta < \pi/4 $,
then study the right half using the Mirror transformation $ {\cal M} $.
The mirror center $ \beta=\pi/4 $ respects the Mirror symmetry.

\section{ Z-x phase and C-IC transition at the low critical field $ h_{c1} $.}
\begin{figure}
\includegraphics[width=7.5cm]{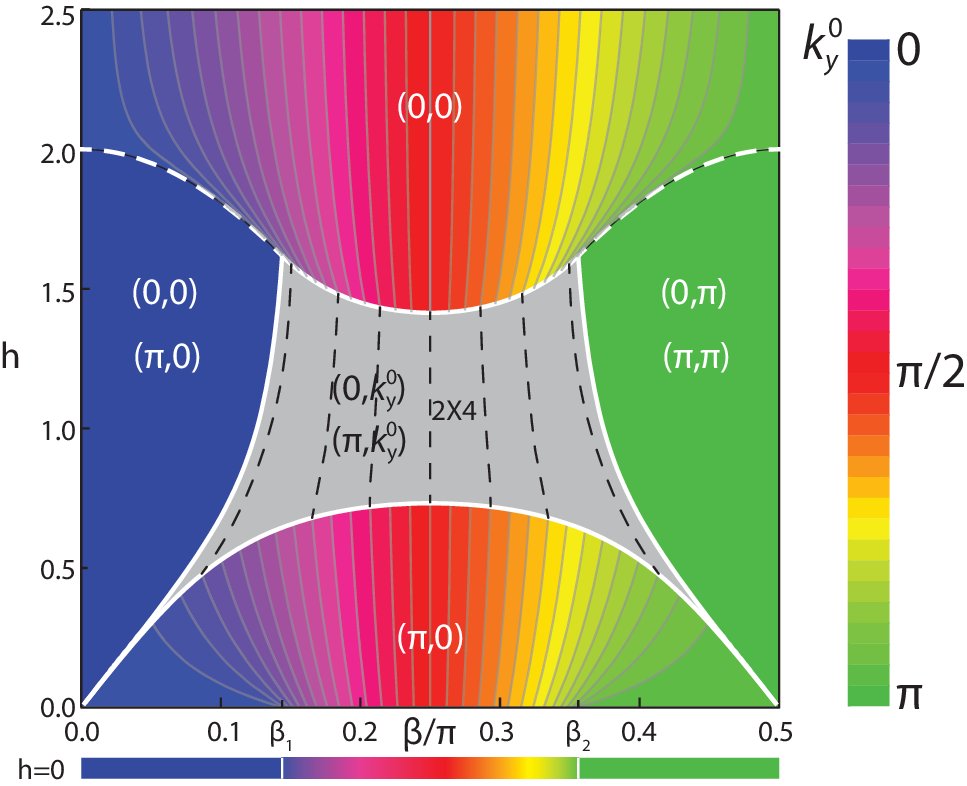}
\caption{  The orbital ordering  wavevectors of  the two collinear, two coplanar and the non-coplanar phases.
The constant contour plot of the minima $ (0, k_y^0 ) $  of the C-IC magnons in the Z-x state at $ h < h_{c1} $ and  Z-FM state
at $ h>h_{c2} $, connected by the orbital ordering wavevectors ( dashed line ) inside the IC-SkX.
There are one C-C transition from the canted phase to the Z-FM at $ ( h_{c2}, 0 < \beta < \beta_1 ) $ with the dynamic exponent $ z=1 $.
There are three different kinds of C-IC transitions at $ h_{c1} $, $ ( h_{c2}, \beta_1 < \beta < \beta_2 ) $ and $ h_L $ ( or $ h_R $ )
from the IC-SkX to the Z-x, Z-FM and canted phase with
the dynamic exponents $ z=2 $, $ z=2 $ and $ (z_x=1, z_y=3 ) $ respectively. }
\label{minima}
\end{figure}

It was shown in \cite{rh} that at $ h=0 $,
the Z-x state is the exact ground state with an excitation gap.
It remains the exact ground state at a small $h$ until the gap closes at $ h_{c1} $.
Any $h> 0$ will turn {\sl all } the C-C$_0$, C-C$_{\pi}$ and C-IC at $h=0$ \cite{rh} into the C-IC magnons
located only at one minimum  ${\bf k}^0=(0, 0< k^{0}_y(\beta,h) < \pi )$ whose constant contour
was shown in Fig.\ref{minima}.
The spin wave spectrum $\omega_{\pm}(\mathbf{k})$ in the reduced Brillioun zone (RBZ) is worked out in the Appendix.
Expanding the lower branch of the spin wave spectrum $\omega_{-}(\mathbf{k})$
near its minimum  $\mathbf{k}= {\bf k}^0 + \vec{q} $ leads to the non-relativistic dispersion:
\begin{align}
	\omega_{Z}(\mathbf{q})=\Delta_Z+\frac{q_x^2}{2m_{Z,x}}+\frac{q_y^2}{2m_{Z,y}}
\label{lowmid}
\end{align}
where $\Delta_Z(\beta,h) $ is the gap
and $ m_{Z,x}(\beta,h), m_{Z,y}(\beta,h)$ are the two effective masses.

The lower critical Zeeman field $h_{c1}$ is determined by $\Delta_Z(\beta,h_{c1} )=0$.
Its expression is given in the appendix and shown in Fig.\ref{globalphase}.
Near $ h \sim h^{-}_{c1} $, $\Delta_Z\sim (h_{c1}-h)^{1}$.
The two effective masses remain non-critical at $ h_{c1} $.
The condensation of the C-IC magnons indicates a transition
from the Z-x state into a IC-SkX phase with the orbital ordering wavevectors  $(0,k^{0}_y)$ ( Fig.\ref{minima} )
which has the dynamic exponent $ z=2 $. The nature of this transition and the IC-SkX phase will be explored further
from $ h_{c1}< h < h_{c2} $ below.


\section{ FM phase, C-C and C-IC transitions at the upper critical field $ h_{c2} $.}
At a strong Zeeman field $ H\gg J>0 $ , the system is in a FM state
subject to quantum fluctuations shown in Fig.\ref{globalphase}.
Its spin wave spectrum $\omega(\vec{k}) $ always has two degenerate minima located at $ (0, k^{0}_{y} ) $ and $ (\pi, k^{0}_{y} ) $ where $ 0 \leq k^{0}_{y} \leq \pi $ shown in Fig.\ref{minima}.
The upper critical field $ h_{c2} $, determined by the vanishing gap at the two minima,
takes a piece-wise form:
\begin{equation}
	h_{c2} =
	\begin{cases}
		1+|\cos2\beta|,	&\beta\in I=(0,\beta_1)\cup(\beta_2,\pi/2)\\
		\sqrt{\frac{3-\cos4\beta}{1-\cos4\beta}}, &\beta\in I\!I=[\beta_1,\beta_2]
	\end{cases}
\label{hc2}
\end{equation}
where the two different pieces indicate transitions to two different states.
At $ \beta= \beta_1, \beta_2 $, the two expressions coincide.
The two values $ \beta_1,\beta_2 $ coincides with the boundaries between $ C-C_0, C-C_{\pi} $ and
C-IC in the $ Z-x $ state at $ h=0 $ at the leading order of linear spin wave expansion (LSWE) in Fig.\ref{globalphase}.
The physical reason for the coincidence is not known.

When $0<\beta<\beta_1$, expanding $\omega(\mathbf{k})$ around $(0,0)$ or $(\pi,0)$ leads to:
\begin{align}
	\omega_{F0}(\mathbf{q})=\sqrt{\Delta^2_{F0}+v_{F,x}^2q_x^2+v_{F,y}^2q_y^2} - c_{F}q_y
\label{highleft}
\end{align}
where $ \Delta_{F0} \sim (h-h_{c_2})^{1/2} $ and $ z=1 $.
At $h=h_{c_2}$, $v_{F,x}=1$, $v_{F,y}=\sqrt{\cos2\beta}$, $c_{F}=\sin2\beta$.
At $\beta=0$, $c_F=0$, Eq.\eqref{highleft} recovers the relativistic form.
The simultaneous condensations of the C-magnons at $\mathbf{k}^0= (0,0)$ and $(\pi,0)$
indicates a transition from the FM state into a canted phase
with the two orbital ordering wavevectors shown in Fig.\ref{minima}.
The nature of the canted phase will be explored further from $ h_{L} < h < h_{c2} $ below.
Increasing $\beta$ along $ h= h_{c_2} $ where $ \Delta_{F0}=0 $,
the slope $v_{F,y}(\beta)-c_F(\beta)$ of the dispersion Eq.\eqref{highleft} at $ q_y >0 $ decreases. At $\beta=\beta_1$, the slope vanishes.
As shown below, the dynamic exponents along $ q_x $ and $ q_y $ directions become anisotropic $ ( z_x=1, z_y=3) $.
This is a multi-critical ( M )  point of
the three phases: FM (collinear), canted (Co-planar) and Incommensurate Skyrmion (IC-SkX) crystal (Non-Coplanar) phase.

When $\beta_1<\beta<\beta_2$, expanding $\omega(\mathbf{k})$ around the two minima $(0,k^{0}_{y}) $ or $(\pi,k^{0}_{y})$,
we obtain a similar non-relativistic form as Eq.\eqref{lowmid}:
\begin{equation}
    \omega_{F}(\mathbf{q}) =\Delta_{F}+\frac{q_x^2}{2m_{F,x}}+\frac{q_y^2}{2m_{F,y}}
\label{highmid}
\end{equation}
where $\Delta_{F}\sim (h-h_{c_2})^1$ and $ z=2 $.
At $ h= h_{c_2} $, $ m_{F,x}=\sqrt{\sin^42\beta-\cos^22\beta}$, $m_{F,y}=\sin^22\beta/m_{F,x}$.
At $ \beta=\beta_1$, $h_{c_2}= 1+\cos2\beta$, $m_{F,x}=0$ and $m_{F,y}=\infty$
which match the anisotropic $(z_x=1,z_y=3)$ dynamic behaviors at the M point in Eq.\eqref{highleft}.
The condensation of the C-IC magnons indicates a transition from the FM state into a IC-SkX phase
with the orbital ordering  wavevectors  $(0,k^{0}_y)$ and $(\pi,k^{0}_y)$ shown in Fig.\ref{minima}.
The spin structure of this IC-SkX phase will be explored further from $h_{c1}< h < h_{c2}$ below.

\section{ Canted phase and C-C transition at $ h_{c2}, 0< \beta < \beta_1 $ and the C-IC transition at the left critical field $ h_L $.}
When $\beta$ is near the Abelian case $\beta=0$,
we first determine the simplest classical state to be a FM state in the XZ plane (Fig.\ref{globalphase}).
Its tilted angle with the $z$ axis is $\theta=\pm\arccos[h/h_{c2}]$
where $ h_{c2}=2\cos^2\beta $ matches the upper critical field achieved
from $ h > h_{c2} $ Eq.\eqref{hc2}.
Note that Eqn.\ref{hc2} is reached from the FM phase by the quantum fluctuations to the order of $1/S $.
The same $ h_{c2} $ is reached from the canted phase below just by the classical minimization. This consistency indicates
that there is a second order transition at $ h_{c2} $ which does not receive a quantum correction at least to the order of $ 1/S $.


\begin{widetext}

\begin{figure}[!htb]
\includegraphics[width=5.8cm]{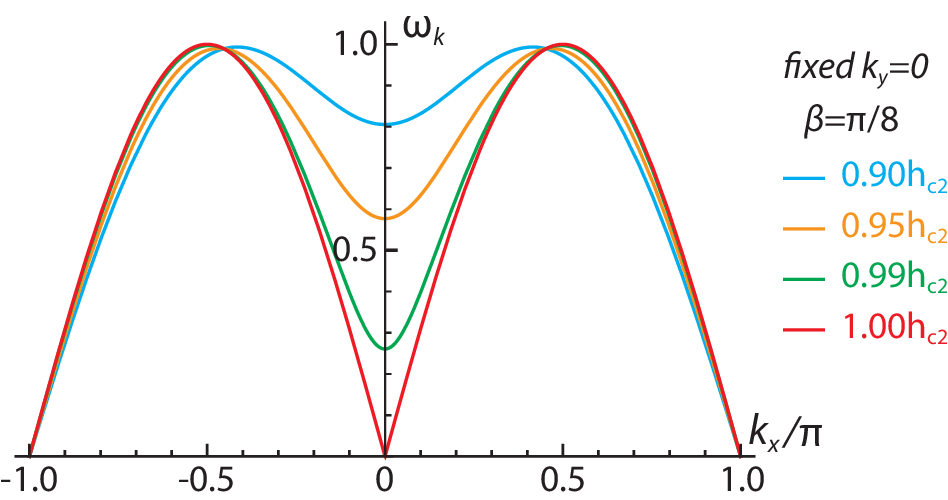}
\quad
\includegraphics[width=5.8cm]{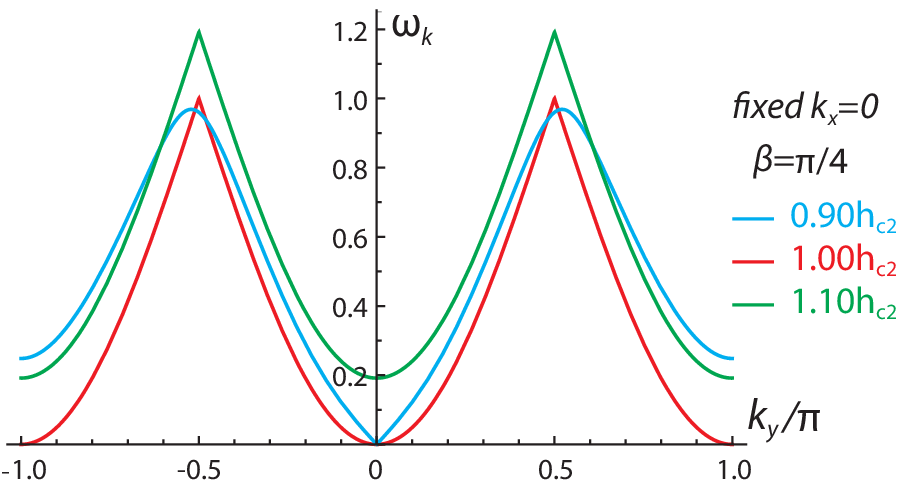}	
\quad
\includegraphics[width=5cm]{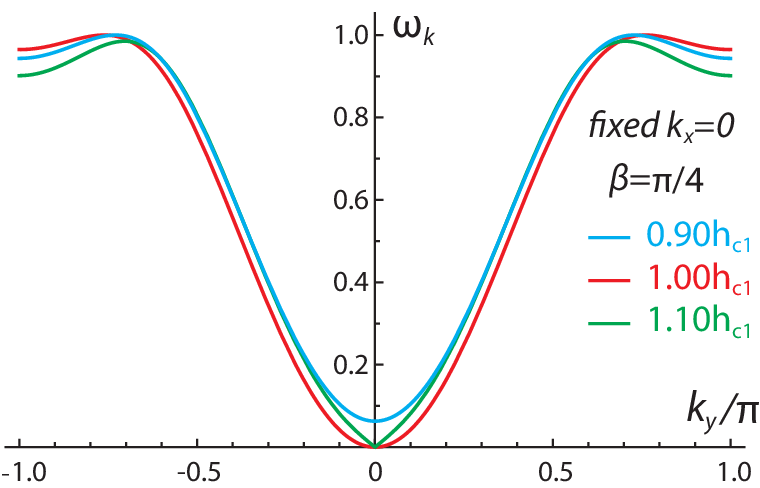}
\caption{ Quantum C-C phase transitions near $ h_{c1} $ and $ h_{c2} $.
          (a) The spin wave dispersion in the canted state as a function of $k_x$ at a fixed $k_y =0 $
          and at $ \beta=\pi/8<\beta_1, h \leq h_{c2}(\beta) $. As $ h \rightarrow h^{-}_{c2} $,
          there is a roton minimum  developing at $ (0,0) $,
          which drives the transition from the canted state to the FM at $ h=h_{c2}(\beta) $ with the dynamic exponent $ z=1 $.
          (b) The spin wave dispersion  $\omega_-(k)$ in the $ 2 \times 4 $ SkX state at $ \beta=\pi/4 $ as a function of $k_y$ with fixed $k_x=0$.
     (b1) Approaching $ h_{c2} $ from below, there is a roton minimum developing at $ (0, \pi) $. From above,
     there are two degenerate minima condensing simultaneously at $ (0,0) $ and  $ (0, \pi) $.
     (b2) Near $h_{c1}$, there is only one minimum dropping at $ (0,0) $ from the Z-x state. Both transitions have the dynamic exponent $ z=2 $. }
\label{fmpolar}
\end{figure}

\end{widetext}

  The most general form of the canted state can be obtained by applying the $ U(1)_{soc} $ symmetry operator \cite{rh}
  $ U_1(\phi)=e^{i \phi \sum_{i} (-1)^{x} S^{z} } $ to the
  FM state in the $ XZ $ plane:
\begin{eqnarray}
   S^{z} & = & \cos \theta
       \nonumber  \\
   S^{+} & = & [S^{-}]^{\dagger}=
  \sin \theta[ \cos \phi + i \sin \phi (-1)^x]
\label{cantclass}
\end{eqnarray}
   It contains the two ordering wavevectors $ \mathbf{Q}_1=(0,0) $
   and $ \mathbf{Q}_2=(\pi,0) $ which match those of the condensed C-magnons coming from $ h> h_{c2} $.
   Setting $ \phi=0, \pi $ recovers the two FM states which does not break the translational symmetry, but breaks
   the $ {\cal P}_z $ and the $ U(1)_{soc} $ symmetry.
  However, when $ \phi \neq 0, \pi $, Eq.\eqref{cantclass} also breaks the translational symmetry.
  For example, setting $ \phi=\pi/2 $ leads to a canted ( co-planar ) state
  $  S^{z}=\cos \theta, S^{+}= \sin \theta e^{i (-1)^x \pi/2 } $ in the $ YZ$ plane which
  breaks the translational symmetry. Naively, if two states break different symmetries of the Hamiltonian, they
  belong to different states. However, here, they belong to the same family of states related by the $ U(1)_{soc} $ symmetry of the
  Hamiltonian. {\sl This counter-intuitive result is a salient feature of the SOC }.

   The $ U(1)_{soc} $ symmetry breaking leads to a gapless Goldstone mode
    $ \phi $ located at $k^0=(\pi,0)$ and takes the rather peculiar form:
\begin{equation}
	\omega_{g}(\vec{q}) =\sqrt{v^2_{g,x} q_x^2+v^2_{g,y} q_y^2}-c_g q_y
\label{gold}
\end{equation}
where $	v_{g,x}, v_{g,y}, c_g=h\tan\beta $ are listed in the appendix.

{\sl (a)  The C-C transition from the canted state to the FM phase at $ h_{c2} $ driven by the Roton dropping. }

Fixing $ 0 < \beta<\beta_1$, as one increases to $h_{c_2}(\beta)$ from below, a roton minimum develops at $(0,0)$
shown in Fig.\ref{fmpolar}a. Its spectrum takes a similar form as Eq.\eqref{highleft}:
\begin{equation}
 \omega_{r}(\vec{k}) =\sqrt{\Delta_r^2+v^2_{r,x} k_x^2+v^2_{r,y}k_y^2}- c_r k_y
\label{roton}
\end{equation}
where $\Delta_r \sim ( h_{c2}- h)^{1/2}, v_{r,x}, v_{r,y},  c_r=h\tan\beta $ are given in the appendix.

At $ \beta=0 $, $ c_g= c_r=0 $, Eq.\eqref{gold} and \eqref{roton} recover the relativistic form.
Along $h=h_{c_2}, \Delta_r=0 $, Eq.\eqref{gold} and \eqref{roton} become the same.
Setting $v_{r,y}-c_r=0 $  leads to the M point at $\beta=\beta_1$.

As shown above, using the FM state in the XZ plane, the Goldstone mode Eqn.\ref{gold} and the roton mode Eqn.\ref{roton} ) are
located at $ (\pi,0) $ and $ (0,0) $ respectively in the full BZ. However, as said above,
choosing $ \phi=\pi/2 $ leads to the canted ( co-planar ) state in the $ YZ $ plane,  then
one need to introduce two HP bosons $a $ and $b $ for the two sublattices A/B respectively,
there are two modes $ \omega_{\pm}(k) $ inside the RBZ.
Both the Goldstone mode $ \omega_{-}( k ) $   and the roton mode $ \omega_{+}( k ) $
are located at $ (0,0) $  in the RBZ.
Obviously, when $ h $ gets close to $ h_{c2} $ from below,
the roton mode $ \omega_{+}( k ) $ becomes degenerate with the Goldstone mode $ \omega_{-}( k ) $, so can not be dropped.

{\sl (b) Bosonic Lifshitz type of C-IC transition from the canted state to the IC-SkX phase at the left critical field $ h_L $. }

At fixed $ h < h_{c2}( \beta_1) $, as the SOC strength increases,
the slope of the Goldstone mode $ v_{g,y}-c_g $ in Eq.\eqref{gold} along $q_y>0$ decreases.
Setting the slope vanishing, we obtain the left critical field $h_L$:
\begin{equation}
	h_L=\frac{2\sin\beta\sqrt{\cos2\beta}}{\sqrt{2+\sec^2\beta-2\sec^4\beta}}.
\label{eq:hL}
\end{equation}
which is shown in Fig.\ref{globalphase}.
Setting $ h_{c2}=h_L $ leads to $\beta=\beta_1$ and $h^*=h_{c_2}(\beta_1)=\frac{\sqrt{5}+1}{2}$ (Golden ratio).
This is the multi-critical ( M ) point
of the three phases: FM (collinear), canted (Co-planar) and IC-SkX (Non-Coplanar) phase.
It has the anisotropic dynamic exponents $ (z_x=1,z_y=3) $.
Near $ h_L $, by expanding the Goldstone mode Eq.\eqref{gold} to higher orders,
 $ \omega(q_x=0,q_y >0 )= (v_{g,y}-c_g)q_y + c_3 q^{3}_y + \cdots $ where $ c_3 >0 $ is given in the appendix.
When $ v_{g,y}-c_g >0 $, the minimum position is at $ q^{0}_y=0 $, so it is in the canted state.
When $ v_{g,y}-c_g <0 $, the minimum position is at $ q^{0}_y= (\frac{c_g-v_{g,y}}{c_3})^{1/2} $,
it is in the IC-SkX state Eqn.\ref{icphase} with the orbital order at $ (\pi,q^{0}_y) $ ( Fig.\ref{minima} ).
Indeed, this infinitesimal small orbital order connects
the one at $ h_{c2}, \beta= \beta^{+}_1 $ smoothly to the one at $ h_{c1}, \beta= 0^{+} $ due to the condensations of C-IC
at $ h_{c2} $ and $ h_{c1} $ respectively.
This is a bosonic type of quantum Lifshitz transition \cite{tqpt,loff}, however, with the odd power of terms such as $ q_y, q^{3}_y,.... $
which is a salient feature due to the SOC. So
it is a completely new class of bosonic type of Lifshitz transition \cite{dimer,dimer1,dimer2}
with the anisotropic dynamic exponent $ (z_x=1,z_y=3) $.

\begin{widetext}

\begin{figure}[!htb]
\includegraphics[width=5.5cm]{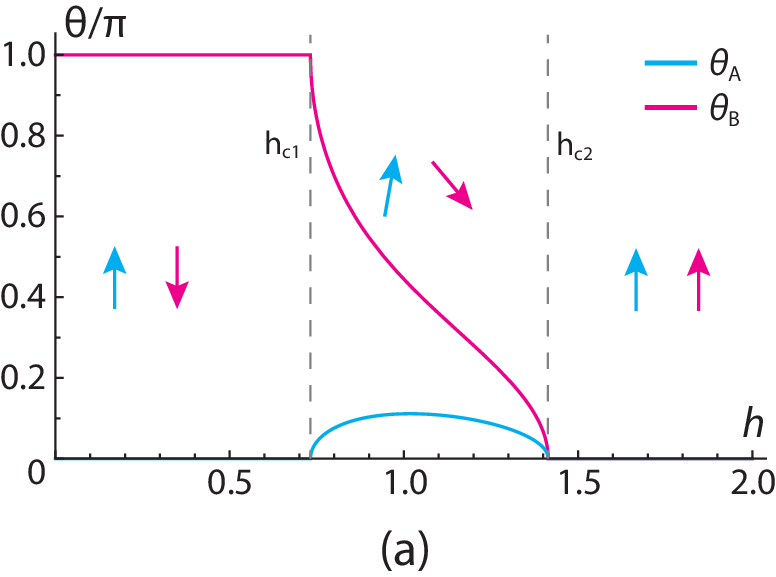}
\quad
\includegraphics[width=5.5cm]{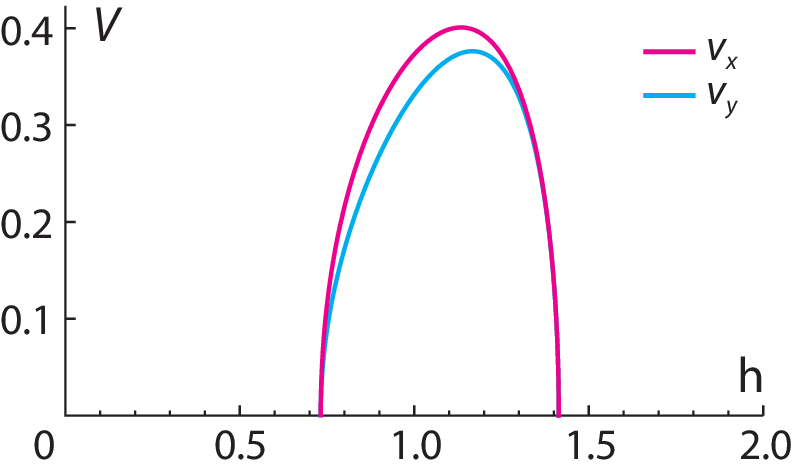}
\quad
\includegraphics[width=5.5cm]{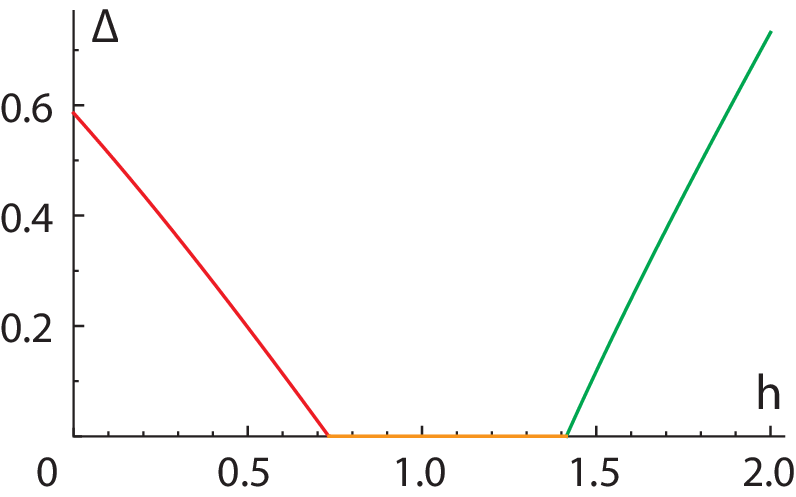}	
	\caption{  The $ 2 \times 4 $ SkX state at $ \beta=\pi/4 $:
(a)  The evolution of the two polar angles  $\theta_A$ and $\theta_B$ in the two sublattices A and B
     as a functions of $h$ from the $ Z-x $ state at $ h < h_{c1} $ to
     the $ 2 \times 4 $ SkX state at $ h_{c1} < h < h_{c2} $, then to the FM state at  $ h > h_{c2} $.
(b) The spin wave velocity $v_{G,x}$, $v_{G,y} $ of the Goldstone mode inside
     the $ 2 \times 4 $ SkX as a function of $h$.
     At $ h_{c1} $ or $ h_{c2} $,  $v_{G,x}= v_{G,y}=0 $, the dispersions become quadratic indicating $ z=2 $.
    (c) The excitation gap $\Delta$ at $ \beta=\pi/4 $ as a function of $h$. It is gapless inside the $ 2 \times 4 $ SkX state.
	    For $ h<h_{c_1} $ and $ h>h_{c_2} $, it is close to be, but not exactly linear.
        Near $ h_{c1}, h_{c2} $, $ \Delta \sim h_{c1}-h $ and $ \Delta \sim h-h_{c2} $ respectively.	}
\label{pi4}
\end{figure}

\end{widetext}

One can also show that $\frac{\partial h_L}{\partial \beta}\Big|_{\beta=0}=
 \frac{\partial h_{c1}}{\partial \beta}\Big|_{\beta=0} = 2 $.
However, $h_L$ is always above $h_{c1}$,
so there is always a narrow window of IC-SkX phase sandwiched between
the collinear $ Z-x $ phase and the co-planar canted phase.
There is NO direct transition between the two.
This is consistent with the contour $ (0, k^{y}_0 \rightarrow 0^{+} ) $
in the $ \beta \rightarrow 0 $ limit from $ h < h_{c1} $ in Fig.\ref{globalphase}.

\section{ Incommensurate Skyrmion crystal (IC-SkX) phases and C-IC transitions at $ h_{c1} $ and $ h_{c2} $.}
As shown in \cite{rh}, due to the explicit $ U(1)_{soc} $ symmetry in the $ U(1)_{soc} $ basis,
the anomalous spin correlation functions in the FM state at $ h > h_{c2} $ vanish.
So the classical state due to the condensation of the IC-magnons
at $(0,k^{0}_{y})$ and $(\pi,k^{0}_{y})$ in Fig.\ref{minima} can be determined in the $U(1)_{soc}$ basis first.
Transforming back to the original basis, then acting on it by the $U_1(\phi)$ leads to the IC-SkX state:
\begin{eqnarray}
   S^{z} & = & A + B (-1)^{x}
       \nonumber  \\
   S^{+} & = & [S^{-}]^{\dagger}=[ C + D (-1)^{x} ] e^{i (-1)^{x} [ k^{0}_{y} y + \phi ] }
\label{icphase}
\end{eqnarray}
which breaks the translational, ${\cal P}_z$ and the $U(1)_{soc}$ symmetry and
has a non-vanishing Skyrmion density  $Q_{ijk}=\mathbf{S}_i\cdot(\mathbf{S}_j\times\mathbf{S}_k)$
where $ i, j, k $ are 3 lattice sites in a square lattice.
However, from Eq.\ref{icphase}, one can see the IC-SkX still keeps the combination of the translation along the $ y $ axis
and $ U(1)_{soc} $: $ y \rightarrow y+1, \phi \rightarrow \phi- k^{0}_{y} $
denoted as $ [U(1)_{soc}]_{\phi \rightarrow \phi- k^{0}_{y} } \times ( y \rightarrow y+1 ) $\cite{socsdw}.
This remaining combined symmetry is important for the calculations of the spin wave spectrum above the IC-SkX phase.
In the following, we will first focus on $\beta=\pi/4$ where $ k^{0}_{y}=\pi/2 $
leads to the commensurate $ 2 \times 4 $ Skyrmion crystal phase shown in Fig.\ref{globalphase}.
Then we will discuss the generic \mbox{IC-SkX} phases when $\beta \neq \pi/4$.

{\sl (a) The commensurate $ 2 \times 4 $ Skyrmion crystal phase at $ k^{0}_{y}=\pi/2 $. }

Minimization of the classical ground state energy leads to the two independent polar angles
$\theta_A$, $\theta_B$ in the two sublattices shown in Fig.\ref{pi4}a.
As shown in the appendix, after making suitable local rotations to align spin quantization axis along the $Z$ axis,
we find the spin wave spectrum $ \omega_{\pm}(\mathbf{k}) $ in the RBZ.
Expanding $ \omega_{-}(\mathbf{k}) $ around the $\Gamma=(0,0) $ point leading to the expected gapless Goldstone mode $ \phi $ :
\begin{equation}
	\omega_{G}(\vec{k})=\sqrt{v_{G,x}^2k_x^2+v_{G,y}^2k_y^2}
\label{gold2}
\end{equation}
where  $ v_{G,x}, v_{G,y} $ are shown in Fig.\ref{pi4}b.

As $ h \rightarrow h^{-}_{c2} $, there is also roton mode developing at $ (0, \pi ) $ which  takes the relativistic form:
\begin{equation}
	\omega_{R}( \mathbf{q} )
	=\sqrt{\Delta^2_R+v_{R,x}^2 q_x^2+v_{R,y}^2 q_y^2},
\label{roton2}
\end{equation}
    where $\mathbf{k}=\mathbf{q}+(0, \pi ) $
and  $ \Delta_R  \sim h_{c2}-h $.

  In fact, as shown in Fig.\ref{pi4}a, putting $ \theta_A= 0, \theta_B=\pi $ and  $ \theta_A= \theta_B=0 $, one can also push the calculations
  to the Z-x state at $ h< h_{c1} $ and the FM state at $ h > h_{c2} $ respectively, of course, in a different gauge than the original one
  used in previous sections. The gaps along the whole central line $ \beta=\pi/4 $ are shown in Fig.\ref{pi4}c.
  As expected, the gaps are gauge invariant, but the minimum positions of excitations may shift at different gauges \cite{rh,tqpt}.
  Indeed, in the original basis, both the Goldstone mode Eq.\eqref{gold2} and the roton mode  Eq.\eqref{roton2} will shift to $ (0,\pi/2 ) $ in the RBZ
  $ -\pi/2 < k_x < \pi/2, -\pi/4 < k_y-\pi/2  < \pi/4 $.

   At $ h_{c_2}(\beta=\pi/4)=\sqrt{2} $, $v_{G,x}=v_{G,y}=0$ and also $ \Delta_R=0$, $v_{R,x}=v_{R,y}=0 $.
   Pushing the expansion to $ k^4 $ in both Eq.\eqref{gold2} and Eq.\eqref{roton2},
   we find the effective masses of both the Goldstone mode and the roton mode
   coincide with the $m_{F,x}$, $m_{F,y}$ achieved from the FM state Eqn.\ref{highmid}, so $ z=2 $.
   Similarly, at $ h_{c_1}(\beta=\pi/4)=\sqrt{3}-1 $, $v_{G,x}=v_{G,y}=0$.
   Pushing the expansion to $ k^4 $ in Eq.\eqref{gold2}, we find the effective masses of the Goldstone mode
   coincide with the $ m_{Z,x}$, $m_{Z,y}$ achieved from the Z-x state Eq.\eqref{lowmid}, so $ z=2 $.
   The critical behaviors at $ h=h_{c2} $ and $ h=h_{c1} $ are shown in Fig.\ref{fmpolar}b1 and \ref{fmpolar}b2 respectively.

{\sl (b) The IC-SkX phase when $ \beta \neq \pi/4 $. }

\begin{figure}
\includegraphics[width=7cm]{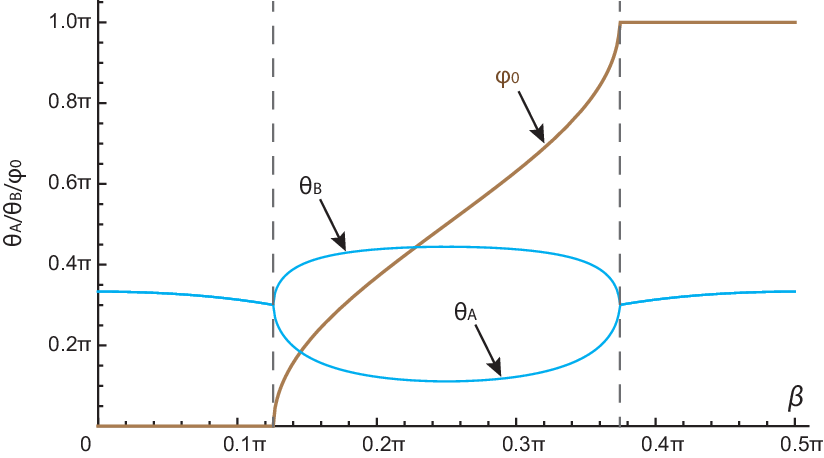}\\
\includegraphics[width=7cm]{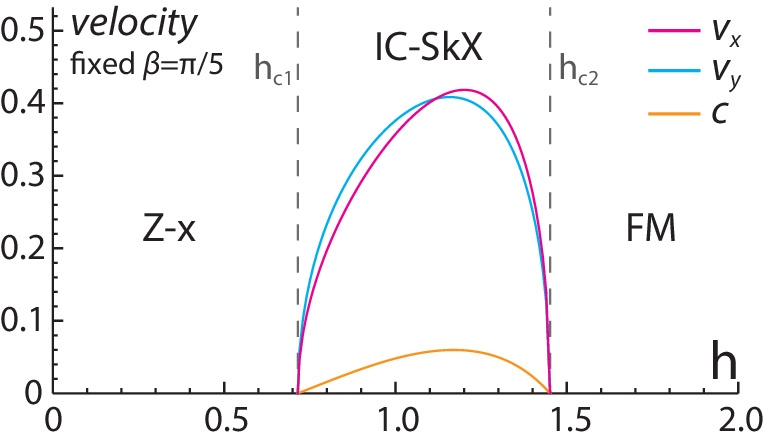}
\caption{ The generic IC-SkX state: (a) At a fixed $ h=1$,  the two angles $\theta_A, \theta_B $ and
the orbital order $ k^{0}_{y} $  of the classical IC-SkX state as functions of $\beta$.
At $ h< h_L $, $ \theta_A=\theta_B $ and $ k^{0}_y=0 $, it is the canted state in the left in Fig.\ref{minima}.
At $ h > h_R $, $ \theta_A=\theta_B $ and $ k^{0}_y=\pi $, it is the canted state in the right in Fig.\ref{minima}.
There is a quantum  Lifshitz type C-IC transition at $ h_L $ ( and $ h_R $ ).
(b) At a fixed $ \beta=\pi/5 $, the three velocities $ v_{G,x}, v_{G,y} $ and $ c_G $ as a function of $ h $.
 At $ h_{c1} $ and  $ h_{c2} $,  $v_{G,x}= v_{G,y}=c_G=0 $, the dispersions become quadratic indicating $ z=2 $.
 There are quantum C-IC transitions at $ h_{c1} $ and  $ h_{c2} $ due to the condensations of C-IC magnons. }
\label{angleandv}
\end{figure}

   The minimization of the classical energy leads to the two independent angles $\theta_A, \theta_B$
   in the two sublattices A and B and also the optimal orbital order $  k^{0}_{y}( \beta, H) $ satisfying $  k^{0}_{y}( \beta=\pi/4, H)=\pi/2 $.
   The results along the horizontal line $ h=1 $  are shown in Fig.\ref{angleandv}a and drawn in Fig.\ref{minima}.

   After making suitably chosen rotations
   to align the spin quantization axis along the $Z$ axis, one need only introduce two HP bosons $a/b$ for the
   two sublattices A/B respectively and perform a Bogoliubov transformation to obtain the spin wave spectrum $ \omega_{\pm}(k) $.
   After lengthy manipulations and very careful long wavelength expansion,
   we find the Goldstone mode $ \phi $ at $ \Gamma=(0,0) $:
\begin{equation}
	\omega_{G}( \vec{k} )=\sqrt{v_{G,x}^2k_x^2+v_{G,y}^2k_y^2}-c_G k_y
\label{goldGc}
\end{equation}
    where $ c_{G} ( \beta, H )= -c_G( \pi/2-\beta, H ) $, so $ c_G > 0 $ when $ \beta < \pi/4 $,
    $ c_G < 0 $ when $ \beta > \pi/4 $ and $ c_G = 0 $ when $ \beta = \pi/4 $ recovering Eqn.\ref{gold2}.
    How the three velocities $ v_{G,x}, v_{G,y} $ and $ c_G $ change from $ h_{c1} $ to $ h_{c2} $ at a fixed $ \beta= \pi/5 < \pi/4 $ are shown in
    Fig.\ref{angleandv}b.

    Similarly, we find a roton mode  developing near $ ( 0, \pi) $ as $ h \rightarrow h^{-}_{c2} $:
\begin{equation}
	\omega_{R}( \vec{q} )
	=\sqrt{\Delta^2_R+v_{R,x}^2 q_x^2+v_{R,y}^2 q_y^2}-c_R q_y
\label{rotonRc}
\end{equation}
    where $ c_{R} ( \beta, H )= -c_R( \pi/2-\beta, H ) $, so $ c_R > 0 $ when $ \beta < \pi/4 $,
    $ c_R < 0 $ when $ \beta > \pi/4 $ and $ c_R = 0 $ when $ \beta = \pi/4 $ recovering Eqn.\ref{roton2}.

    Comparing with Eqn.\ref{gold} and \ref{roton}, we find the Goldstone mode and the Roton mode in the IC-SkX phase
    take similar forms as those in the canted phase.
    At a fixed $ h $ in Fig.\ref{minima}, we find that as $ h \rightarrow h^{+}_{L} $
    ( or $ h \rightarrow h^{-}_{R} $ ) , $ v_{G,y}- c_G \rightarrow 0 $ ( or $ v_{G,y}+ c_G \rightarrow 0 $ ),
    it is a bosonic Lifshitz transition with the anisotropic dynamic exponent $ z_x=1, z_y=3 $.
    This picture is completely consistent as that achieved from the canted phase to the IC-SkX.
    These facts suggest some sort of duality between the cant phase and the IC-SkX phase on the two side of $ h_L $ ( or $ h_R $ ) in Fig.\ref{minima}.

    Taking $ h \rightarrow h^{+}_{c1} $, $ v_{G,x}=v_{G,y}=0 $ and $ c_G=0 $ in Eqn.\ref{goldGc}, expanding it to the order $ k^{4} $, we find it
    matches Eqn.\ref{lowmid} reached from $ Z-x $ state below $ h_{c1} $.
    Taking $ h \rightarrow h^{-}_{c2} $, $ v_{G,x}=v_{G,y}=0 $ and $ c_G=0 $ in Eqn.\ref{goldGc} and
    $ \Delta_R=0, v_{R,x}=v_{R,y}=0 $ and $ c_R=0 $ in Eqn.\ref{rotonRc}, expanding both equations to order $ k^{4} $, we find both
    matches $ m_{F,x} $ and $ m_{F,y} $ in Eqn.\ref{highmid} reached from FM state above $ h_{c2} $.

\section{ Mirror Reflection  about $ \beta=\pi/4 $. }
  One can define a Mirror transformation $ {\cal M } $ which consists the local rotation
  $ \vec{\tilde{S}}_{i} =R(\hat{x},\pi ) R(\hat{z},\pi n_2) \vec{S}_{i}$ followed by a Time reversal transformation $ {\cal T } $.
  Under  $ {\cal M } $, $ (\beta, h ) \rightarrow ( \pi/2 - \beta, h ) $.
  Only $ \beta=\pi/4 $ is invariant under $ {\cal M } $.
  Note that this anti-unitary Mirror transformation is defined in SOC parameter space instead of position space.

  When $ 0 < h < h_{c1} $, under  $ {\cal M } $, the gap minimum $  (0, k^{0}_y ) $  at $ (\beta, h ) $ is mapped to
  the gap minimum  $  (0, \pi- k^{0}_y ) $ at $ ( \pi/2 - \beta, h ) $.
  This mapping also applies to $ h > h_{c2} $ where there is one extra minimum at $ (\pi, k^{0}_y  ) $.
  So the identity $ k^{0}_y +  (\pi-k^{0}_y)= \pi $  explains the reflection symmetry in the minimum positions about $ \beta=\pi/4 $
  in the $ ( \beta, H ) $ plane in Fig.\ref{globalphase} and Fig.\ref{minima}.
  When $ h_{L} <  h < h_{c2} $ in the left hand side of Fig.\ref{minima}, the two minima  at $ (0,0) $ and $ (\pi,0) $
  are mirror reflected to those at $ (0,\pi) $ and $ (\pi,\pi) $ when $ h_{R} <  h < h_{c2}  $ in the right hand side.
  The right critical field $ h_R $ is given by Eqn.\ref{eq:hL} by setting $ \beta \rightarrow \pi/2 - \beta $ which is the mirror reflected image
  with respect to $ \beta= \pi/4 $ in Fig.\ref{globalphase}.

  For any state $ |\psi \rangle_L $ on the left $ \beta \leq \pi/4 $, one can get the state on the right by the mirror transformation
  $ |\psi \rangle_R = {\cal M} |\psi \rangle_L  $ where  $ {\cal M}={\cal T} e^{i \frac{\pi}{2} \sigma_x} e^{i \frac{\pi}{2} y \sigma_z} $.
  Indeed, applying the operation on the canted state Eqn.\ref{cantclass} on the left side
  leads to the canted state on the right hand side:
\begin{eqnarray}
     S^{z} & = &  \cos \theta,   \nonumber  \\
  S^{+} &  = &   [S^{-}]^{\dagger}= -\sin \theta (-1)^{y} e^{-i (-1)^x \phi }
\end{eqnarray}
  which contains two ordering wavevectors $ \vec{Q}_1=(0,\pi) $  and $ \vec{Q}_2=(\pi,\pi) $ shown in Fig.\ref{minima}.
  Setting $ \phi=\pi $ gives the state shown on the right in Fig.\ref{globalphase}.
  Applying $ {\cal M} $ on the IC-SkX state Eqn.\ref{icphase} leads to $ k^{0}_y \rightarrow \pi- k^{0}_y $.
  Applying it on the state $ 2 \times 4 $ SkX state at $ k^{0}_y =\pi/2 $ leads back to itself as expected.

\section{  Finite Temperature properties  }

 Any experiments are performed at finite temperatures which are controlled by the quantum phases and phase transitions at $ T=0 $ in Fig.1 and Fig.2.
 Here, we discuss the effects of finite temperatures.

{\sl 1. Physical quantities at Low temperatures }

\begin{widetext}

\begin{figure}[!htb]
\includegraphics[width=10cm]{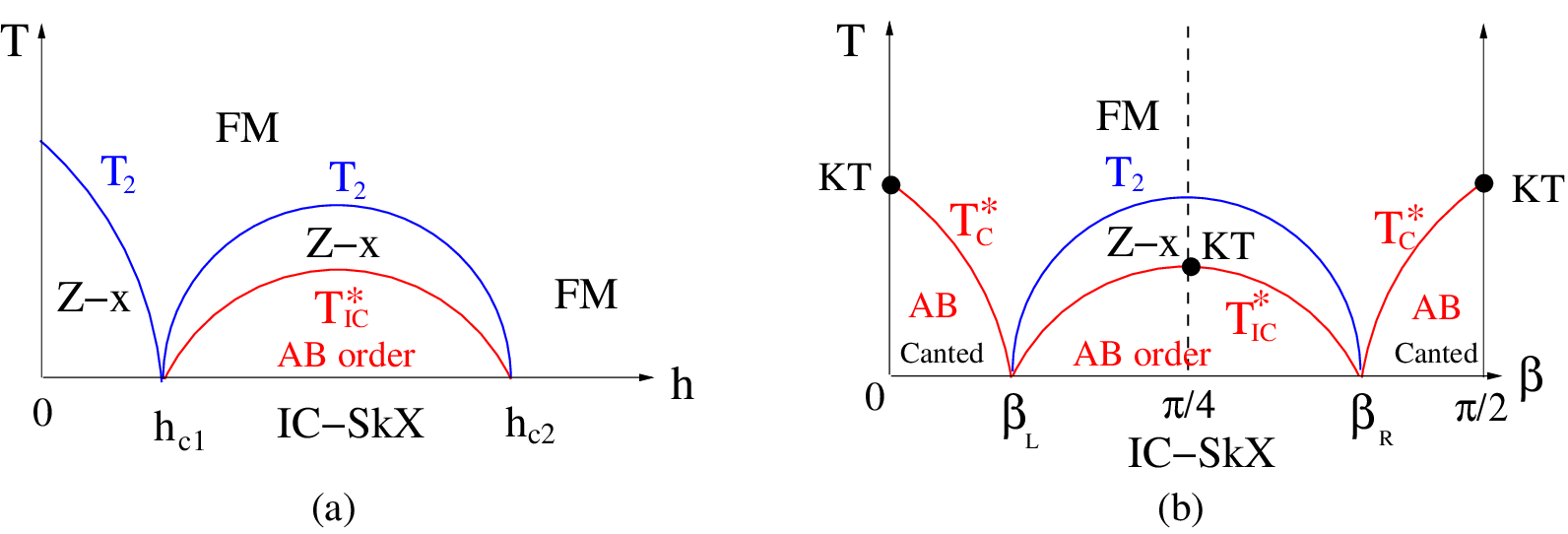}
\hspace{0.5cm}
\includegraphics[width=4.5cm]{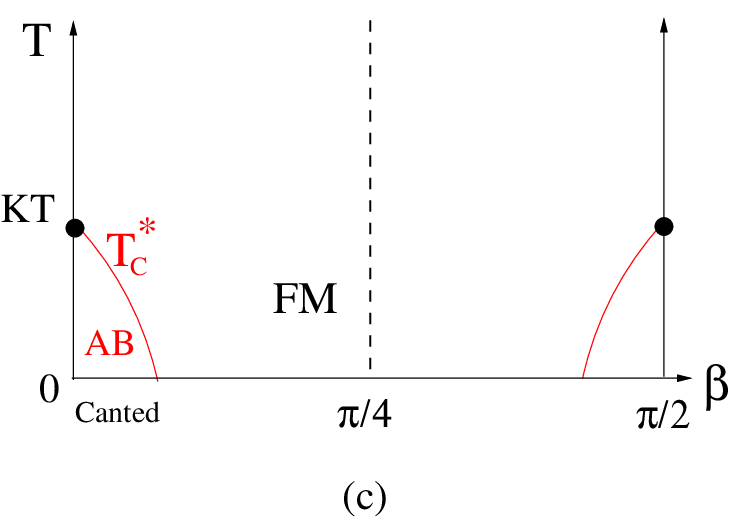}
\caption{  (Color online) Finite temperature phase transitions above the three quantum C-IC transition at $ T=0 $.
(a) At a fixed $ \beta $. At $ T=0 $, there is a quantum C-IC transition  from the Z-x to the IC-SkX at $ h=h_{c1} $ and
from the IC-SkX to the FM at $ h=h_{c2} $ shown in Fig.1. There a finite temperature Ising transition $ T_2 $ above the Z-x state.
The IC-SkX has only an algebraic  ( denoted as AB in the figure ) order in the transverse spin components before
getting to the $ Z-x $ state at $ T =T^{*}_{IC} $, then melt into the FM state at $ T_2 $.
In general, the $ T^{*}_{IC} $ may be  different from the Koterlitz-Thouless (KT) transition except at at the mirror symmetric point $ \beta=\pi/4 $
as shown as the black dot in (b).
(b) At a fixed $ h $.  At $ T=0 $, there is a quantum C-IC transition  from the canted phase to the IC-SkX at $ \beta_L $ and
from the IC-SkX to the mirror reflected canted phase at $ \beta_R=\pi/2-\beta_L $ shown in Fig.1.
There is a finite temperature $ T^{*}_C $ transition above the canted state in the same universality class as the $ T^{*}_{IC} $.
Only at the two Abelian points $ \beta=0, \pi/2 $, the $ T^{*}_{IC} $ reduces to the KT transition.
There is a mirror symmetry about $ \beta=\pi/4 $ where the IC-SkX reduces to the $ 2 \times 4 $ SkX and the $ T^{*}_{IC} $ reduces to the KT transition.
Replacing the IC-SkX in (b) by the FM leads to (c) where there is a C-C transition  from the canted to the FM state at $ T=0 $ in Fig.1.
As argued in Sec.VIII-a, all the critical temperatures $ T_c \sim  \Delta \sim 2S J  = NJ \sim  N \times 0.2 nK  $ where the $ N $ is the number of atoms per site, so all the critical temperatures can be easily increased above the experimentally reachable temperatures simply by increasing the number of spinor atoms
on every lattice site. }
\label{htzx}
\end{figure}

\end{widetext}

   Following \cite{rh}, one can work out the thermodynamic quantities such as magnetization, uniform and staggered susceptibilities, specific heat
  and Wilson ratio at the low temperatures in all the 5 phases in Fig.\ref{globalphase}.
  For example, the specific heat in the $ Z-x $ state at $ h< h_{c1} $ and the FM state at $ h> h_{c2} $
  take the same form as that at the $ h=0 $  achieved in Ref.\cite{rh} by just using the two
  $ (\beta, h ) $ dependent effective masses $ m_x= m_x(\beta, h), m_y=m_y(\beta,h) $.
  Of course, due to the Goldstone modes in the canted and the IC-SkX phases, the specific heat in the two phases takes the power law $ C_v \sim T^2 $.
  Similarly, one can work out various kinds of spin correlation functions at the low  temperatures.
  Following the procedures \cite{scaling,tqpt}, one can also derive the scaling functions of these physical quantities
  at finite temperatures across the three C-IC quantum transitions in Fig.\ref{htzx}a,b and also the C-C transition from the canted phase to the FM at the left
  or right segment of $ h_{c2} $ in Fig.\ref{htzx}c.


{\sl 2. Finite temperature phase transitions }

   As argued in \cite{rh}, there is only one finite temperature phase transition in the Ising universality class \cite{rhtran} above the $ Z-x $  phase.
   The FM state breaks no symmetries of the Hamiltonian, so no transitions above it. So we only need to discuss
   the finite temperature transitions above the canted phase and IC-SkX state as shown in Fig.\ref{htzx}a.

{\sl (a) The canted phases: }
   In the canted phase, from Eq.7, one can see that at any $ T>0 $, the Goldstone mode fluctuations Eq.8 lead to $ \langle S^{+} \rangle=0 $,
   so the transverse spin correlation functions display algebraic orders at the two ordering wavevectors $ \vec{Q}_1=(0,0) $ and $ \vec{Q}_2=(\pi,0) $.
   So there is only one finite temperature phase transition $ T^{*}_{C} $ driven by the topological defects
   in the phase $ \phi $ in Eq.\ref{cantclass} above the canted phase to destroy the algebraic order ( Fig.\ref{htzx}b,c ).
   In view of the rather peculiar anisotropic form of the Goldstone mode in Eq.8,
   we expect that the $ T^{*}_{C} $ belongs to a new universality class different than
   the conventional  Kosterlize-Thouless (KT) one.

   The transverse Bragg spectroscopy in the canted phase at $ T=0 $ will display sharp peaks at  $ \vec{Q}_1=(0,0) $ and $ \vec{Q}_2=(\pi,0) $.
   However at $ 0 < T < T^{*}_{C} $, the transverse peaks at  $ \vec{Q}_1 $ and $ \vec{Q}_2 $ will be  replaced by some power law singularities \cite{socsdw}.
   At $  T > T^{*}_{C} $, the power law singularities disappear.

{\sl (b) The IC-SkX phase: }
   In the IC-SkX phase, from Eq.11, one can see that at any $ T>0 $, the Goldstone mode fluctuations Eqn.1
   also lead to $ \langle S^{+} \rangle=0 $,
   so the transverse spin correlation functions also display algebraic orders at the four
   in-commensurate ordering wavevectors  $ (0, \pm k^{0}_y ) $ and $ (\pi, \pm k^{0}_y ) $.
   So there are two finite temperature phase transitions above the IC-SkX state:
   one transition $ T^{*}_{IC} $ in the transverse spin sector to destroy the algebraic order, then
   another Ising  $ Z_2 $  transition in the longitudinal spin sector $ T_2 $ to destroy the $ A $ and $ B $ sublattice $ Z_2 $
   symmetry breaking as shown in Fig.\ref{htzx}a.
   In view of the similar form of Eq.15 as Eq.8, we expect the $ T^{*}_{IC} $ is in the same universality class as  $ T^{*}_{C} $
   above the canted phase.
   We also expect $ T^{*}_{IC} < T_{2} $. Of course, at all the quantum phase transition boundaries in Fig.1, $ T^{*}_{IC}=T_{2}=0 $.



   The elastic longitudinal Bragg spectroscopy in the IC-SkX at $ T=0 $ will display a sharp peak at  $ (\pi,0) $,
   while the transverse Bragg spectroscopy will display sharp peaks at  the four
   in-commensurate ordering wavevectors  $ (0, \pm k^{0}_y ) $ and $ (\pi, \pm k^{0}_y ) $.
   However at $ 0 < T < T^{*}_{IC} $, the transverse peaks at  $ (0, \pm k^{0}_y ) $ and $ (\pi, \pm k^{0}_y ) $ will be
   replaced by  some power law singularities \cite{socsdw},
   the longitudinal peak remains sharp. At $ T^{*}_{IC} < T < T_{2} $, the power law singularities disappear, but the longitudinal peak remains sharp.
   When $ T > T_2 $, the longitudinal peak disappears.

{\sl 3. High temperature expansion  }

   When the temperature is well above the critical temperatures of all the quantum phases in Fig.1,
   we may perform a high temperature expansion which is complementary to the spin wave expansion at temperyares well below
   the critical temperatures in {\sl 1} above.
   Following \cite{rh}, when $ T \gg J, h_{c2} $, we may also perform a high temperature expansion in terms  of $ J/T $ and $ h/T $
   where mixing terms in $ J/T $ and $ h/T $ are expected.
   The connections between the Wilson loop and specific heat can also be established.

\section{ Implications on recent cold atom experiments and materials with SOC. }

  We will discuss the implications on cold atoms and SOC materials respectively.
  The two experimental systems have different advantages and limitations to
  explore different aspects of the rich many body phenomena in Fig.1 and Fig.2.

{\sl (a) Experimental realizations and detections in the original and the $ U(1)_{soc} $ basis in cold atoms.}

In the original basis Eq.\eqref{rhh}, the gauge field configuration is achieved by putting
$ \pi/2 \sigma_x $ in the $x$-bond, $ \beta \sigma_z $ in the $y$-bond and
the Raman laser induced Zeeman field $ H $ along the $ \hat{z} $ direction.
The two  Abelian points $ \beta=0, \pi/2 $ have been realized in previous experiments
\cite{stagg1,stagg2,stagg3,stagg4,uniform1,uniform2,uniform3,newexp,newexp2,sorev}.
As pointed out in \cite{uniform3},
the $ \beta \sigma_z $ in the $y$-bond can be achieved by adding spin-flip Raman lasers
or by driving the spin-flip transition with RF or microwave fields.

As discussed in \cite{tqpt}, one of the big advantages of cold atom experiments over condensed matter systems is that different gauges
can be realized in cold atoms, so both gauge non-invariant and gauge invariant  quantities can be measured in cold atom experiments.
As shown in \cite{rh}, the $ U(1)_{soc} $ basis
$ \tilde{\mathbf{S}}_n = R( \hat{x}, \pi n_1 ) \mathbf{S}_n $
may be more easily realized experimentally.
In the $ U(1)_{soc} $ basis, the RFHM in the Zeeman field Eq.\eqref{rhh} becomes $
   H_{U(1)_{soc},h}=H_{U(1)_{soc}}- H \sum_i (-1)^x S^z_{i} $
where $ H_{U(1)_{soc}} $ is the RFHM in the $ U(1)_{soc} $ basis at $ H=0 $ given in \cite{rh} and
the Zeeman field becomes a staggered one along $ \hat{x} $ direction.
Then  applying the $ R(\hat{x}, n_1 \pi) $ on
   all the states shown in Fig.\ref{globalphase} leads to the corresponding states in the $ U(1)_{soc} $ basis. The thermodynamic quantities are gauge invariant, so
   are the same in both basis.
   But the spin-correlation functions are gauge dependent, need to be re-evaluated in the $ U(1)_{soc} $ basis at both low and high temperatures.

Recently, using the optical Raman lattice scheme, the authors in the experiment \cite{2dsocbec} indeed realized the SOC
with tunable $ (\alpha, \beta) $ in a square lattice and the direction and magnitude
of the Zeeman field $ \vec{H} $ are tunable. An optical lattice clock scheme \cite{clock} was proposed to
suppress the heatings issue and generate a 2d SOC in an optical lattice.
Most recently, by using the most magnetic fermionic element dysprosium to eliminate the heating due to the spontaneous emission,
the authors in \cite{ben} created a long-lived SOC gas of quantum degenerate atoms. The long lifetime of this weakly interacting SOC degenerate
Fermi gas will facilitate the experimental study of quantum many-body phenomena manifest at longer time scales,
So the novel phases and phase transitions in Fig.\ref{globalphase} and \ref{minima} are ready to be
explored in near future cold atom experiments.

  As  noted in \cite{rh} and repeated at the very beginning, the RFH model Eq.\ref{rhgeneral} is for spin $ S=N/2 $ where $ N $ is the number of atoms per site.
  As estimated in \cite{rh},  taking some typical values of cold atoms in the strong coupling limit,
  $ t \sim 3 $nK, $ U \sim 50 $nK, the critical temperatures in Fig.\ref{htzx} at 2d would be $ T_c \sim J \sim t^2/U \sim
  0.2 $nK  for a spin $ S=1/2 $ RFH at $ U>0 $. It remains experimentally quite challenging to reach such a low temperature.
  However, because the critical temperature scales as $ T_c/J \sim 2S $, so
  if even taking $ S=5 $, then  $ T_c \sim 2 nK $.
  In view of new cooling techniques
  \cite{cool1,cool2} to reach $ 0.35 $nK, this enhanced critical temperature $ T_c $ should
  be reachable with the current cold atom experimental cooling techniques \cite{rafm2}.
  In fact, the $ T_c $ can be enhanced further by going to a cubic lattice,
  but with no SOC along the $ \hat{z} $ direction. Adding the $ \hat{z} $ direction without putting the SOC along it will not
  increase the experimental difficulties \cite{2dsocbec}, but will certainly increase the critical temperatures.
  In fact, there have been extensive experimental efforts to investigate the AFM correlations \cite{coldafm} in SOC free fermionic systems.
  However, the AFM is for spin $ S=1/2$ and gapless, so $ T_c=0 $ at 2d.
  It is a remarkable property of the $ S=N/2 $ RFH: its suppressed critical temperature $ T_c \sim J \sim t^2/U $ can be
  compensated by increasing the atom number $ N $ per site. Unfortunately, the RAFH may not share such a nice properties \cite{rafm,rafm2}.
  So in the aspect of temperature requirements, it would be easier to study the IC-SkX correlations in Fig.1 and Fig.2
  than to study the AFM correlations \cite{rafm2}.

 As argued in \cite{rh}, all the physical quantities calculated in Sec.VII-1 can be precisely
 determined by various experimental techniques  such as dynamic or elastic, energy or momentum resolved, longitudinal or
 transverse atom or light Bragg spectroscopies \cite{lightatom1,braggbog,braggangle,braggeng,braggsingle,braggsoc},
 specific heat measurements \cite{heat1,heat2}  and {\sl In-Situ} measurements \cite{dosexp}.

{\sl (b) Implications to materials with strong SOC }

  Although the RFHM was derived as the strong coupling model of interacting spinor boson Hubbard model at integer fillings in the presence of SOC,
  we may just treat it as an effective lattice quantum spin model which incorporate competitions among
  AFM Heisenberg  physics, FM Kitaev physics and  DM physics. The Zeeman field adds a new dimension to these competitions.
  So RFHM + H can be used to not only to describe cold atom systems, but also the universal features of some strongly
  correlated materials which host some of these interactions.

The IC-SkX phase in Fig.\ref{globalphase} can be realized in some materials with a strong Dzyaloshinskii-Moriya (DM) interaction \cite{dm1}.
Indeed, a 2D skyrmion lattice has been observed between $ h_{c1}=50$ mT  and $ h_{c2}=70$ mT
in some chiral magnets \cite{sky4} MnSi or a thin film of Fe$_{0.5}$Co$_{0.5}$Si \cite{sky4}.

The  3d hyperhoneycomb  iridates  $\alpha,\beta,\gamma$-Li$_2$IrO$_3$ was previously considered to be a promising candidate to
realize Kitaev spin liquid phases. Unfortunately, so far, no sign of any spin liquids was detected in this so called Kitaev materials.
Instead, an incommensurate, counter-rotating (in A/B sublattice), non-coplanar magnetic orders
with the ordering wavevector $ \vec{q}=(0,0,q), q= \pi + \delta, \delta \sim 0.14 \pi $ lying along the orthorhombic $ \vec{a} $ axis
was detected on the iridates \cite{kitaevlattice,kitaevlattice1,kitaevlattice2}.
Most remarkably, the IC-SkX phase Eq.\eqref{icphase} is strikingly  similar to the this state.
In the following, we provides some insights and explanations on the Magnetic orderings in Iridates $ \alpha,\beta,\gamma-Li_2 Ir O_3 $
from the RFHM+H perspective.

As shown in \cite{rh}, when expanding the two $R$ matrices in Eqn.\ref{rhgeneral},  one can see thatit leads to
a Heisenberg + Kitaev  ( or quantum compass model in a square lattice \cite{comp} ) + Dzyaloshinskii-Moriya (DM) interaction $
 H_{s}=-J [\sum_{\langle i j \rangle  } J^a_{H} \vec{S}_{i} \cdot \vec{S}_{j}
	+\sum_{\langle i j \rangle a } J^{a}_{K} S^{a}_{i} S^{a}_{j}
	+\sum_{\langle i j \rangle a } J^{a}_{D} \hat{a} \cdot \vec{S}_{i} \times \vec{S}_{j} ]
$
where $ \hat{a}= \hat{x}, \hat{y} $, $ J^{x}_H=\cos 2 \alpha,  J^{y}_H=\cos 2 \beta $;
$ J^{x}_K= 2 \sin^2 \alpha,  J^{y}_K= 2 \sin^2 \beta $ and $ J^{x}_D=\sin 2 \alpha,  J^{y}_D=\sin 2 \beta $.


Obviously,  at $ \alpha=\beta =0 $, the Hamiltonian becomes the usual FM Heisenberg model
$ H= -J \sum_{ \langle i j \rangle } \vec{S}_{i} \cdot  \vec{S}_{j} $.
At one end of the solvable line $ (\alpha=\pi/2, \beta=0) $,
we get the FM Heisenberg model in one rotated basis
$ H= -J \sum_{\langle ij\rangle } \vec{\tilde{S}}_{i} \cdot \vec{\tilde{S}}_{j} $,
where the  $\vec{\tilde{S}}_{i} = R(\hat{x},\pi n_1) \vec{S}_{i}$.
At the other end of the solvable line $ (\alpha=\pi/2, \beta=\pi/2) $,
we get the FM Heisenberg model in another rotated basis
$ H= -J \sum_{\langle ij\rangle } \vec{\tilde{S}}_{i} \cdot \vec{\tilde{S}}_{j} $,
where $ \vec{\tilde{S}}_{i} = R(\hat{x},\pi n_1)  R(\hat{y},\pi n_2) \vec{S}_{i} $.

Along the whole solvable line  $ (\alpha=\pi/2, \beta ) $, we can write:
$ J^{x}_H=-1,  J^{y}_H=\cos 2 \beta;~~~~~
  J^{x}_K= 2,  J^{y}_K= 2 \sin^2 \beta;~~~~~
  J^{x}_D=0,  J^{y}_D=\sin 2 \beta  $.
  It is easy to see $  J^{y}_H > 0 $ when $ \beta < \pi/4 $, $  J^{y}_H < 0 $ when $ \beta > \pi/4 $ and vanishes at $ \beta=\pi/4 $.
  While $  J^{y}_K= 1 $ and $ J^{y}_D=1 $ at  $ \beta=\pi/4 $.
  Obviously, the FM Kitaev term dominates, plus a AFM Heisenberg term in both bond when $ \beta > \pi/4 $, plus a DM term in XZ plane
  $  J \sin 2 \beta ( S_{ix}S_{jz}- S_{iz}S_{jx}) $. In the presence of the Zeeman term $ H $, the $ RFHM+H $
  leads to the IC-SkX state in the center regime in Fig.1 and Fig.2. The IC-SkX matches well the
  the incommensurate, counter-rotating (in A/B sublattice), non-coplanar magnetic orders detected by neutron
  and X-ray diffractions on iridates \cite{kitaevlattice,kitaevlattice1,kitaevlattice2} $ \alpha,\beta,\gamma$-Li$_2$IrO$_3$.
  So our $ RFHM-H  $ could be an alternative to the minimal $ (J,K,I) $ model used in \cite{kitaevlattice,kitaevlattice1,kitaevlattice2}
  or to the minimal $ (J,K, \Gamma ) $ model used in \cite{kim,kim1} to fit the experimental data phenomenologically.
  Of course, both the $ (J,K,I) $ and $ (J,K, \Gamma ) $ model were directly extracted from the spin, orbital and crystal structures
  of the material itself.
  One common thing among all the three models is that it is dominated by FM Kitaev term, plus a small AFM Heisenberg term.
  However, our $ RFHM+H $ has only two independent parameters $ ( \beta, H) $.
  We reach the global phase diagram Fig.1 in $ ( \beta, H) $ by the well controlled $ 1/S $ quantum fluctuations calculations.
  We also achieved the magnon spectra Eqn.\ref{goldGc}, \ref{rotonRc} above the IC-SkX phase.
  While the solutions of the minimal $ (J,K,I) $ model have to involve un-controlled Luttinger-Tisza approximation,
  those of minimal $ (J,K, \Gamma ) $ model involve analytical or numerical calculations only at the classical level.

  These SOC materials are automatically in the strong coupling regime.
  So far, there is no experimental data on the magnon spectrum above the IC-SkX phase in the Iridates $ \alpha,\beta,\gamma-Li_2 Ir O_3 $.
  However, in contrast to the cold atom systems in (a), its total angular momentum $ J=1/2 $ is fixed due to the crystal field splitting,
  it is difficult to tune various parameters to study the three classes of quantum C-IC transitions in Fig.1 and Fig.2.



\section{ Discussions }
    The classical Commensurate to In-commensurate (C-IC) transitions are discussed in the context of
    adatom adsorption on periodic substrates such as graphite \cite{tom,dual1}.
    However, it seems there are very little works on quantum in-commensurate phases and associated quantum  C-IC transitions.
    There are previous theoretical works on In-commensurate spin density waves ( IC-SDW) in the $ J_1-J_2-J_3 $ frustrated
    quantum Heisenberg model \cite{j1j2j3}.
    The in-elastic neutron scattering experiments \cite{Csdw} on the high $ T_c $ cuprate  $ La_{2-x} Sr_{x} Cu O_4 $
    indeed found that the magnetic peak at momentum $ ( \pi,\pi) $ in the AFM state near half filling splits into four incommensurate peaks at
    $ ( \pi \pm \delta, \pi \pm \delta ) $ in the underdoped and superconducting regime. The
    incommensurability $ \delta $ scales as the doping concentration $ x $.
    It was known that this IC-SDW is collinear and is due to the geometric frustrations in the quantum Heisenberg model with the spin $ SU(2) $ symmetry.
    Our theoretical work discovered that it is the combination of the SOC and the Zeeman field in a bipartite lattice which leads to the IC-SkX state in a large
    parameter space in Fig.1 and Fig.2.
    The IC-SkX is non-coplanar with non-vanishing Skyrmion density instead of collinear.
    So the geometric frustrations and the SOC are two completely different mechanisms leading to the  in-commensurate phases
    which also own very different properties in the two cases.

     Usually, there could just be a direct second order transition between two compatible commensurate phases.
     However, between any two in-compatible commensurate phases, there can only be four possible routes:
     (1) A direct first order transition (2) through some in-commensurate phases (3) through a quantum spin liquid phase
     (4) through a de-confined quantum critical point.
     Our main results well crafted in Fig.\ref{globalphase} and Fig.\ref{minima} show that  the case (2) is happening here for the RFHM in a Zeeman field.
     There are one quantum C-C transition from the Z-FM state to the canted state at $ ( h_{c2}, 0 < \beta < \beta_1 ) $.
     Most importantly, there are 3  quantum C-IC transitions: the Z-x state to the IC-SkX at $ h_{c1} $,
     the Z-FM state to the IC-SkX at $ ( h_{c2}, \beta_1 < \beta < \beta_2 ) $,
     the canted state to the the IC-SkX at $ h_L $  ( or $ h_R $ ).
     The first two are due to the the condensation of C-IC magnons driven by the external Zeeman field $ h $, while the third is due to
     the bosonic quantum Lifshitz transition driven by the SOC. All are second order quantum phase transitions, but in different universality classes.
     It is the interplay between the SOC and the Zeeman field which leads to the
     spin-orbital correlated  collinear, co-planar ( canted), non-coplanar ( Skyrmion crystal ) phases
     in a square ( which is a bipartite ) lattice.
     Fig.\ref{globalphase} and Fig.\ref{minima} can be contrasted with
     the collinear magnetic phases in a bipartite lattice, spiral or non-coplanar magnetic phases found in geometrically frustrated lattices \cite{sachdev}.
     The effective actions  and renormalization group analysis of all the quantum phase transitions ( especially the quantum C-IC transitions ) with the dynamic
     exponents $ z=1 $, $ z=2 $ and the anisotropic ones $ (z_x=1, z_y=3) $ in Fig.1 will be studied,
     the finite temperature transitions above all the 5 phases will be investigated,
     the RFHM + H model at generic SOC parameters $ (\alpha, \beta) $ will be presented in separate publications.


 It is instructive to compare the $ ( z_x=1,z_y=3 ) $  bosonic Lifshitz C-IC transition at $h=h_{L}$, $h_R$ in
 Fig.\ref{globalphase} with the quantum dimer model (QDM)
used to describe the transition from one valence bond solid (VBS) to another VBS state with possible intervening in-commensurate VBS \cite{dimer,dimer1,dimer2}.
It was known that near the solvable Rokhsar-Kivelson (RK) point, in the height representation,
the transition can be described by a low energy effective quantum bosonic Lifshitz action
$ {\cal L}_{RK}= \frac{1}{2} [ (\partial_\tau h)^2 + K_2 ( \nabla h )^2 + K_4 ( \nabla h )^4 + \cdots ] $
where $ K_2=0 $ at the RK point with the dynamic exponent $ z=2 $.
It maybe interesting to explore the possible interesting connections between
the $ ( z_x=1,z_y=3 ) $ bosonic Lifshitz C-IC transition at $ h=h_L $, the multi-critical ( M ) point and the IC-SkX phases in Fig.\ref{globalphase}
with the $ z=2 $
bosonic Lifshitz transition, the RK point and the incommensurate tilted VBS phases in the QDM.
However, the degree of freedoms in the QDM is dimers instead of quantum spins.
Our model is a quantum spin model in the presence of both SOC and Zeeman field, so maybe more experimentally accessible than the QDM.

     In short, quantum spin systems with SOC subject to a Zeeman field opens a new platform to
     display rich and novel class of quantum commensurate (C) and In-commensurate (IC) phases, excitations and
     quantum  C-C and C-IC phase transitions, which can be observed in
     both cold atoms and materials with SOC.
     The results achieved in this paper just reveals a tip of an iceberg.

{\bf Note added: } During the review process of this manuscript, based on the proposal to use internal atomic states as effective
" synthetic dimensions " \cite{clock}, the fermionic optical lattice clock scheme
was just successfully implemented for both $ ^{87} Sr $ clock in \cite{clock1}
and $ ^{173} Yb $ clock in \cite{clock2}, where  the heating and atom loss from spontaneous emissions are eliminated, the exceptionally long lifetime
$ \sim 100 s $ of the excited clock state have been achieved.
The " synthetic dimensions " idea has also been used \cite{SDRb} to generate strong SOC  in an effective
two-dimensional manifold of discrete atomic momentum states of $ ^{87} Rb $.
As advocated by all the three experimental groups \cite{clock1,clock2,SDRb}, these ground-breaking experiments set-up a
very promising platform to observe novel many-body phenomena due to
interplay between SOC and interaction in optical lattices. They also open up a new frontier of combining clock precision measurement,
metrology and many body phenomena unique to SOC. It is quite promising that the novel many body phenomena in Fig.1 and Fig.2 could be observed in
these cold atom experiments also in the near future.

{\bf Acknowledgements}

We thank Shuai Chen, Youjin Deng, W. Ketterle, Ruquan Wang and Jing Zhang for helpful discussions on current and future experimental status.
We acknowledge  NSF-DMR-1161497 and AFOSR FA9550-16-1-0412 for supports. The work at KITP was supported by NSF PHY11-25915.
W.M. Liu is supported by NSFC under Grants No. 10934010 and No. 60978019, the NKBRSFC under Grants No. 2012CB821300.

\appendix

\begin{widetext}

In this appendix, we provide some technical details on the results achieved in the main text:
(1) the $ Z-x $ state below $ h_{c1} $, (2) the FM state above $ h_{c2} $, (3) the canted state on the left side $ h_L < h < h_{c2} $,
(4) the incommensurate Skyrmion ( IC-SkX ) states at generic $ \beta $ which reduces to the
$ 2 \times 4 $ SkX state at $ \beta= \pi/4 $.

\section{ Lower critical field $ h_{c1} $ in the $ Z-x $ state }

Following \cite{rh}, we perform SWE on Eqn.M1 to leading order in $ 1/S $.
Introducing the two Holstein-Primakoff (HP) bosons for the two $A/B$ sublattices respectively and
introducing a unitary transformation:
\begin{equation}
	\begin{pmatrix}
		a_k\\
		b_k\\
	\end{pmatrix}
	=
	\begin{pmatrix}
		\sin\frac{\theta_{k,h}}{2} &\cos\frac{\theta_{k,h}}{2}\\
		-\cos\frac{\theta_{k,h}}{2}&\sin\frac{\theta_{k,h}}{2}\\
	\end{pmatrix}
	\begin{pmatrix}
		\alpha_k\\
		\beta_k\\
	\end{pmatrix}
\end{equation}
    where the matrix  elements are given by:
\begin{equation}
	\sin\theta_{k,h}
	=\frac{\cos k_x}{\sqrt{\cos^2 k_x+(\sin2\beta\sin k_y-h)^2}},~~~
	\cos \theta_{k,h}
	=\frac{\sin2\beta\sin k_y-h}{\sqrt{\cos^2 k_x+(\sin2\beta\sin k_y-h)^2}}
\end{equation}
    Setting $ h=0 $ reduces to the unitary transformation in \cite{rh}.

     One can put the Hamiltonian into the diagonal form :
\begin{equation}
	\mathcal{H}_L=E_0+4JS\sum_k[\omega_{+}(k)\alpha_k^\dagger\alpha_k+\omega_{-}(k)\beta_k^\dagger\beta_k]
\label{lowfield}
\end{equation}
where $ E_0= -2NJS^2 $ is the ground state energy ( the same as that at $ h=0 $ ),
$ \vec{k} $ belongs to the reduced Brillioun zone (RBZ) and the spin wave spectrum is:
\begin{equation}
	\omega_{\pm}(k)=1-\frac{1}{2}\cos2\beta\cos k_y \pm\frac{1}{2}\sqrt{\cos^2 k_x+(\sin 2\beta\sin k_y-h)^2}
\label{zxs}
\end{equation}
   whose minima location $ (0, k_y^0) $ is one of the roots of following quartic equation
\begin{equation}
	\sin^2 2\beta \sin^4 k_y
	-2h\sin 2\beta \sin^3 k_y
	+(1+h^2-\sin^22\beta-\sin^42\beta)\sin^2 k_y
	+2h\sin^3 2\beta \sin k_y
	-h^2\sin^22\beta
	=0
\end{equation}

  It turns out that there is always one and only one physical root. The constant contour of  $ k_y^0 $ is shown in Fig.2.

  Expanding $\omega_{-}(k)$ near the minimum  $\vec{k}= {\bf k}^0 + \vec{q} $ leads to:
\begin{equation}
	\omega_{Z}( \vec{q} )=\Delta_Z+\frac{q_x^2}{2m_{Z,x}}+\frac{q_y^2}{2m_{Z,y}}
\label{lowmida}
\end{equation}
  where the  $ \Delta_Z(\beta,h)$ is the gap of the $ C-IC $ magnons
  and $ m_{Z,x}(\beta,h), m_{Z,y}=m_{Z,y}(\beta,h)$ are their two effective masses.
  By comparing with the expansion $ \omega_{Z}( \vec{q} )=\Delta_Z+ c q^z + \cdots $, one can identify the dynamic exponent $ z=2 $.

  The lower critical magnetic field $ h_{c1} $ is determined by $ \Delta_Z(\beta,h_{c1} )=0 $ and is given by:
\begin{align}
	h^8\cos^4 2\beta
	-2h^6(3-10\sin^2 2\beta+6\sin^4 2\beta+\sin^6 2\beta)
	 -h^4(15+36\sin^2 2\beta-31\sin^4 2\beta-28\sin^6 2\beta-\sin^8 2\beta)
\nonumber\\
	  -2h^2(4-29\sin^2 2\beta+6\sin^4 2\beta+50\sin^6 2\beta+5\sin^8 2\beta)
	   -\sin^2 2\beta(8+\sin^2 2\beta)(1-3\sin^2 2\beta)^2=0
\label{hc1}
\end{align}
When $\beta=\pi/4$, it simplifies to:
\begin{align}
	9h^4-72h^2-36=0
	\Longrightarrow
	h_{c_1}=\sqrt{3}-1
\end{align}

  Expanding $\Delta_Z$ around $h_{c1}$, we obtain
\begin{equation}
	\Delta_Z=\frac{\partial \Delta_Z}{\partial h}\Big|_{h=h_{c1}}(h-h_{c1})
	=\frac{\sin2\beta\sin k_y^0-h_{c1}}{2(2-\cos2\beta\cos k_y^0)}(h-h_{c1})
\end{equation}

  The coefficient is nonzero for $\beta\neq0,\pi/2$, thus we obtain $\Delta\sim (h_{c1}-h)^{1}$ whose slope is given in Fig.\ref{twomass}a.
  The values of the two effective masses at $ h_{c1} $ are shown in Fig.\ref{twomass}b.

  At $ \beta=\pi/4 $, the minimum is at $ k^{0}_y= \pi/2 $, expanding around the minimum  $k=q+(0,\pi/2)$ leads to:
\begin{equation}
	\omega_{Z}( \vec{q} ) =2-\sqrt{2+2h+h^2}
	+\frac{q_x^2}{2\sqrt{2+2h+h^2}}
	+\frac{(1+h)q_y^2}{2\sqrt{2+2h+h^2}}
\label{modezx}
\end{equation}
   which gives the dynamic exponent $ z=2 $  and the critical mass at $ h_{c1}= \sqrt{3}-1 $ shown
   in Fig.\ref{twomass}b: $m_{Z,x}=\sqrt{2+2h_{c_1}+h_{c_1}^2}=2$ and $m_{Z,y}=\sqrt{2+2h_{c_1}+h_{c_1}^2}/(1+h_{c_1})=2/\sqrt{3}$ which will be used to compare with those achieved from  $ h_{c1} < h < h_{c2} $ in Sec.IV.



\begin{figure}
\includegraphics[width=13cm]{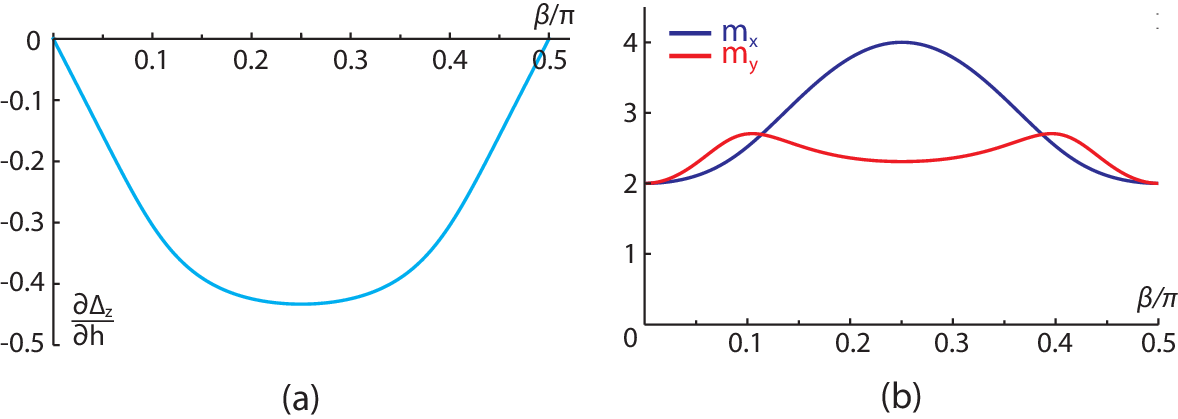}
\caption{ The $ Z-x $ state at $ h < h_{c1} $.
(a) Critical slope $\partial\Delta_Z/\partial h$ as a function of $\beta$
(b) The two effective masses $ m_{Z,x}(\beta,h_{c1})$ and $m_{Z,y}(\beta,h_{c1})$ at the lower critical field $ h_{c1} $ as a functions of $\beta$. }
\label{twomass}
\end{figure}

\section{ Determinations of three segments of $ h_{c2} $ from high field FM state }

   The crucial difference of the upper critical field $ h_{c2} $ from the lower critical field $ h_{c1} $ is that one has to split $ h_{c2} $ into
   3 different segments ( namely, piece-wise ) shown in Fig.1.
   It indicates transitions to 3 different class of states: two canted states and one IC-SkX state.

  The FM state in the high  field $ h > h_{c2} $  breaks no symmetry of the Hamiltonian.
  The $ U(1)_{soc} $ symmetry dictates there must be at least two degenerate minima in the excitations above the FM state.
  So one only need to introduce one HP boson. After performing a Bogliubov transformation, we obtain:
\begin{equation}
	\mathcal{H}_H = -NH(S+\frac{1}{2})+JS\sum_k \omega_k
	+2JS\sum_k \omega_k\alpha_k^\dagger \alpha_k
\label{highfield}
\end{equation}
  where the spin wave dispersion is
\begin{equation}
	\omega_k=\sqrt{(h-\cos2\beta\cos k_y)^2-\cos^2 k_x}-\sin2\beta\sin k_y
\label{fms}
\end{equation}

   Due to the reflection symmetry about $ \beta=\pi/4 $, we only need to focus on $0<\beta<\pi/4$.
   It is easy to see that as dictated by the $ U(1)_{soc} $ symmetry, there are always two degenerate minima located at $k_x=0,\pi$.
   The minimization in $k_y$ leads to:
\begin{equation}
	0=\frac{\partial \omega_k}{\partial k_y}
	=\frac{\cos2\beta\sin k_y(h-\cos2\beta\cos k_y)}
	{\sqrt{(h-\cos2\beta\cos k_y)^2-1}}
	-\sin2\beta\cos k_y
\label{eq:dE/dy}
\end{equation}
    If Eq.\eqref{eq:dE/dy} has a real solution $k_y^0$, then plugging it back into Eq.\eqref{fms} leads to:
\begin{equation}
	\omega_{\rm min}=\frac{\sin k_y^0(h\cos2\beta-\cos k_y^0)}{\sin2\beta\cos k_y^0}
\end{equation}
   The gap vanishing condition is:
\begin{equation}
	\cos k_y^0=h\cos2\beta
\label{eq:Gap0}
\end{equation}

   In fact, Eq.\eqref{eq:dE/dy} is a quartic equation of $\cos k_y$
\begin{equation}
	\sin^22\beta\cos^2 k_y+
	(\cos^2 2\beta-\cos^2 k_y)
	(h-\cos2\beta\cos k_y)^2=0
\label{eq:Quartic}
\end{equation}

If there exists one root with $ \cos k^{0}_y \leq 1 $ in Eq.\eqref{eq:Quartic},
substituting Eq.\eqref{eq:Gap0} into Eq.\eqref{eq:Quartic} leads to
\begin{equation}
	\frac{1}{8}h_{c_2}^2\sin^2 4\beta
	[3-h_{c_2}^2+ ( h_{c_2}^2-1 )\cos4\beta]=0
	\Longrightarrow
	h_{c_2}=\sqrt{\frac{3-\cos4\beta}{1-\cos4\beta}}
\end{equation}

If all positive roots of Eq.\eqref{eq:Quartic} require $\cos k_y^0>1$,
then the minimum is located at $\cos k_y^0=1$, namely, $k^0_y=0$.
Then substituting $\cos k_y^0=1$ into Eq.\eqref{fms} leads to:
\begin{equation}
	0=\omega_{\min}=\sqrt{(h_{c_2}-\cos2\beta)^2-1}
	\Longrightarrow
	h_{c_2}=1+\cos 2\beta
\end{equation}
  which is also the condition ensuring  a real spectrum.

   Combining the two piece-wise $h_{c_2}$ equations leads to:
\begin{equation}
	1+\cos 2\beta=\sqrt{\frac{3-\cos4\beta}{1-\cos4\beta}}
	\Longrightarrow \beta=\beta_1
\end{equation}
   which is shown in Fig.1.

   After extending to $\beta\in(0,\pi/2)$, we obtain:
\begin{equation}
	h_{c2} =   \left \{ \begin{array}{ll}
    1+|\cos2\beta|,  & ~~~ \beta\in I=(0,\beta_1)\cup(\beta_2,\pi/2)  \\
    \sqrt{\frac{3-\cos4\beta}{1-\cos4\beta}},  & ~~~ \beta\in II=[\beta_1,\beta_2]
    \end{array}     \right.
\label{hc2a}
\end{equation}

  The two different piece-wise forms of $ h_{c2} $ in the regime I and II indicates transitions to two different states: canted state
  and IC-SkX state respectively with the dynamic exponents $ z=1 $ and $ z=2 $ respectively.

  At $ \beta=\pi/4, k^{0}_y=\pi/2 $, expanding around the two minima $ (0,\pi/2) $ or  $(\pi,\pi/2) $, Eqn.\ref{fms} becomes:
\begin{equation}
	\omega_{F1}( \vec{q} )=\sqrt{h^2-1}-1+\frac{q_x^2}{2\sqrt{h^2-1}}
	+\frac{q_y^2}{2},
\label{modefm}
\end{equation}
   where $ \vec{k}=(0,\pi/2) + \vec{q} $ or $ \vec{k}=(\pi,\pi/2) + \vec{q} $. It
   gives the two masses at $ h_{c2}=\sqrt{2}$: $m_{F,x}=\sqrt{h_{c_2}^2-1}=1, m_{F,y}=1$ and $ z=2 $ which will be compared to
   those achieved from $ h <h_{c2} $.

\section{ The Goldstone and Roton mode in the canted state }

    It is most convenient to perform SWE on the simplest FM state in the $ XZ $ plane with $ \phi=0 $ in Eqn.M5.
  We first make a global rotation $R_y(\theta)$ to align the spin quantized axis along  the $Z$ axis
  then only need to introduce one HP boson to perform the SWE $ \mathcal{H}=\mathcal{H}_0+\mathcal{H}_1 +\mathcal{H}_2+\cdots $.
  We obtain $ \mathcal{H}_0 = -2NJS^2\Big[\cos^2\beta+\frac{h^2}{4\cos^2\beta}\Big] $,
  $ \mathcal{H}_1=0 $ which is dictated by the correct classical ground state Eqn.M5 and
\begin{equation}
	\mathcal{H}_C  =  -2NJS\cos^2\beta+JS\sum_k\omega_k
	+2JS\sum_k\omega_k\alpha_k^\dagger\alpha_k
\label{leftcant}
\end{equation}
  where the spin wave excitation spectrum is
\begin{equation}
  \omega_k=\sqrt{A_k^2-B_k^2}-C_k
\end{equation}
   where
\begin{eqnarray}
	A_k&=&2\cos^2\beta+\Big(1-\frac{h^2}{4\cos^4\beta}\Big)\cos k_x
	+\Big(\frac{h^2\sin^2\beta}{4\cos^4\beta}-\cos^2\beta\Big)\cos k_y,  \nonumber \\
	B_k&=&\frac{h^2}{4\cos^4\beta}\cos k_x
	+\sin^2\beta\Big(1-\frac{h^2}{4\cos^4\beta}\Big)\cos k_y,  \nonumber \\
	C_k&=& h\tan\beta\sin k_y.
\end{eqnarray}
    The excitation spectrum is always gapless at $ {\bf k^0}=(\pi,0) $.

    Expanding around $ \vec{k}= {\bf k^0} + \vec{q} $, we obtain the Goldstone mode Eqn.M6
\begin{equation}
	\omega_g( \vec{q} ) =\sqrt{v^2_{g,x} q_x^2+v^2_{g,y} q_y^2}-c_g q_y
\label{lowleft}
\end{equation}
    where
\begin{equation}
    v^2_{g,x}=\Big(\frac{h^2(1+\sin^2\beta)}{4\cos^4\beta}-\sin^2\beta\Big),~~
	v^2_{g,y}=\Big(\frac{h^2(1+\sin^2\beta)}{4\cos^4\beta}-\sin^2\beta\Big)\cos2\beta,~~
	c_g=h\tan\beta
\label{golda}
\end{equation}

    As $ h $ increases to $ h_{c2}, 0 < \beta < \beta_1 $, there is also a roton minimum
    developing at  $ (0,0) $ shown in Fig.3a:
\begin{equation}
     \omega_r( \vec{k} ) =\sqrt{\Delta_r^2+v^2_{r,x} k_x^2+v^2_{r,y}k_y^2}- c_r k_y
\end{equation}
    where
\begin{eqnarray}
	\Delta_r^2&= & 4\cos^2\beta-\frac{h^2}{\cos^2\beta}
	=4[1-(h/h_{c_2})^2]\cos^2\beta, ~~~~~ c_r =h\tan\beta.     \nonumber  \\
	v^2_{r,x} &= & \frac{h^2(2+\cos^2\beta)}{4\cos^4\beta}-\cos^2\beta-1,~~~~
	v^2_{r,y} =1+\cos2\beta\cos^2\beta+\frac{h^2(\cos2\beta\sin^2\beta-1)}{4\cos^4\beta},
\label{rotona}
\end{eqnarray}

   At $ h=h_{c2}=1+ \cos 2\beta $,  we find the Goldstone mode at $ (\pi,0) $ and
   the roton mode at $ (0,0) $ achieved from below  $ h \leq h_{c2}, 0 < \beta < \beta_1 $ coincide with those achieved from above
   $ h \geq h_{c2}, 0 < \beta < \beta_1 $, namely: $  v_{g,x}=v_{r,x}=v_{F,x}=1,
   v_{g,y}=v_{r,y}=v_{F,y}=\sqrt{\cos 2 \beta } $ and $ c_g=c_r=c_F=\sin 2\beta $.
   This indicates the transition from the FM to the canted state maybe a second order transition with $ z=1 $.

   Now we look at the Bosonic Lifshitz type of transition at $ h=h_L $ from the canted state to the IC-SkX state.
   We need to perform higher-order gradient expansion around $(\pi,0)$ in the canted state near $ h=h_{L} $
   to see the nature of the transition:
\begin{equation}
	\omega_g( \vec{q} )
	=\sqrt{v_{g,x}^2 q_x^2+v_{g,y}^2 q_y^2
	+v_{xx} q_x^4+v_{yy} q_y^4+v_{xy}q_x^2q_y^2}
	-c_g q_y+c'q_y^3
\label{lifshitz}
\end{equation}
where
\begin{eqnarray}
	v_{xx}
	& = & \frac{1}{4}
	 -\frac{h^2(7+\sin^2\beta)}{48\cos^4\beta}
	  +\frac{1}{12}\sin^2\beta,
\quad
	v_{yy}
	=\Big(\frac{1}{4}
	 -\frac{h^2(1+7\sin^2\beta)}{48\cos^4\beta}
	  +\frac{1}{12}\sin^2\beta\Big)\cos2\beta,
\nonumber \\
	v_{xy} & = & \frac{1}{2}\Big(1-\frac{h^2}{4\cos^4\beta}\Big)\cos^2\beta,
\quad
	c'=(h/6)\tan\beta
\end{eqnarray}
Note that $v_{xx}$ is not positive-define, but it will not cause any instabilities due to $v_{g,x}>0$.
Both $v_{g,y}$ and $v_{yy}$ are positive.

 Without losing the physics of the transition from the canted state to the IC-SkX, setting $q_x=0$ simplifies Eqn.\ref{lifshitz} to:
\begin{equation}
	\omega_g(q_x=0,q_y)=\sqrt{v_{g,y}^2q_y^2+v_{yy}q_y^4}-c_g q_y+c'q_y^3
	=v_{g,y}|q_y|+\frac{v_{yy}}{2v_{g,y}}|q_y|^3-c_g q_y+c'q_y^3
\end{equation}

  Because the transition first happens in $ q_y >0 $ direction, one can see that when $ v_{g,y}-c_g >0 $,
  the minimum position is at $ q^{0}_y=0 $, so it is in the canted state.
  When $ v_{g,y}-c_g < 0 $,
  the minimum position is at $ q^{0}_y= ( \frac{c_g-v_{g,y}}{c_3} )^{1/2} $ where $ c_3= \frac{v_{yy}}{2v_{g,y}} + c^{\prime} $, then
  it is in the IC-SkX state with the orbital order at $ (\pi,  q^{0}_y ) $. Note that the IC-SkX states breaks the $ U(1)_{soc} $ symmetry,
  so acting on the state with the orbital order $ (\pi,  q^{0}_y ) $ will generate a generic IC-SkX state with the two orbital orders
  $ (0,  q^{0}_y ) $ and $ (\pi,  q^{0}_y ) $ shown in Fig.1a.
  Indeed, this infinitesimal small orbital order connects
  the one at $ h_{c2}, \beta= \beta^{+}_1 $ smoothly to the one at $ h_{c1}, \beta= 0^{+} $.
  This is a bosonic type of Lifshitz transition, however, with the odd power of terms such as $ q_y, q^{3}_y,.... $  which is due to the SOC.
  So it is completely new class of bosonic type of Lifshitz transition with the anisotropic dynamic exponents $ (z_x=1, z_y=3) $.

\section{ The Goldstone and Roton mode in the IC-SkX states }

 We follow the procedures used in the main text. We first discuss the two modes in the  $ 2 \times 4 $ commensurate SkX at $ \beta =\pi/4 $.
 Then we extend the calculations to the generic  IC-SkX states when  $ \beta \neq \pi/4 $.

\subsection{ $ 2 \times 4 $ commensurate SkX at $ \beta =\pi/4 $ }

    At $ \beta =\pi/4 $, putting $ k^{0}_y= \pi/4 $ in Eqn.M11 leads to the explicit form of the $ 2 \times 4 $ SkX state:
\begin{eqnarray}
	S_i &=&S(\sin\theta_A\cos(\phi_A-i_y\pi/2),\sin\theta_A\sin(\phi_A-i_y\pi/2),\cos\theta_A),
	~i\in A;   \nonumber \\
	S_j &=&S(\sin\theta_B\cos(\phi_B+j_y\pi/2),\sin\theta_B\sin(\phi_B+j_y\pi/2),\cos\theta_B),
	j\in B;
\label{icphasea}
\end{eqnarray}

    Using Eqn.\ref{icphase}, we find the classic ground state energy  as
\begin{equation}
	E_{\rm classic}=
	-NJS^2[\cos^2\theta_A-\cos\theta_A\cos\theta_B
	+\cos(\phi_A+\phi_B)\sin\theta_A\sin\theta_B
	+h(\cos\theta_A+\cos\theta_B)]
\label{classpi4}
\end{equation}
Because $\sin\theta_A\sin\theta_B>0$, the minimization requires $\phi_A+\phi_B=0$.
For $\theta_A$ and $\theta_B$, the minimization leads to:
\begin{eqnarray}
	&&-\frac{1}{NJS^2}\frac{\partial E_{\rm classic}}{\partial\theta_A}
	=-\sin2\theta_A+\sin(\theta_A+\theta_B)-h\sin\theta_A=0, \nonumber  \\
	& &-\frac{1}{NJS^2}\frac{\partial E_{\rm classic}}{\partial\theta_B}
	=\sin(\theta_A+\theta_B)-h\sin\theta_B=0.
\label{saddleSkX}
\end{eqnarray}
    whose solutions lead to the two polar angles shown in Fig.4a.

    We have also taken a general $4\times4$ structure,
    then numerically minimize the classic ground state energy
    with respect to the 32 parameters (16 polar angles + 16 azimuthal angles).
    Numerical results always find the configuration shown in Fig.4a.

    In Eqn.\ref{icphase}, setting $ \phi_A=-\phi_B= \phi $  which is nothing but the Goldstone mode,
    Eqn.\ref{icphase} can be written in the form of Eqn.M11. We find  \cite{skxdensity}
    an oscillating Skyrmion density $ Q_{ijk}= \vec{S}_i \cdot ( \vec{S}_j \times \vec{S}_k )=
    \frac{(-1)^y}{2} \sin \theta_A \sin 2 \theta_B ( \cos \phi - \sin \phi)- \cos \theta_A \sin^2 \theta_B $
    where $ i, j,k $ are taken as three lattice points around a square.
    At $ \phi=0 $, it reduces to $ (-1)^y \sin ( \theta_A- (-1)^y \theta_B) \sin \theta_B $


   From Eqn.\ref{icphase}, we make suitably chosen rotations $ \tilde{S}_i=R_y(\theta_i)R_z(\phi_i) S_i $ where
   $ \phi_i=\phi-i_y\pi/2,~\theta_i=\theta_A, i \in A; \phi_j=-\phi+j_y\pi/2,~\theta_j=\theta_B,j\in B $
   to align the spin quantization axis along the $Z$ axis. Then one need only introduce two HP bosons $a/b$ for the
   two sublattices A/B respectively and perform a Bogoliubov transformation to obtain:
\begin{equation}
   \mathcal{H}_S=E_0 +2JS\sum_k[ \omega_{+}(k)\alpha_k^\dagger\alpha_k
	+\omega_{-}(k)\beta_k^\dagger\beta_k]
\label{middle}
\end{equation}
withe the spin wave spectrum
\begin{equation}
 \omega_{\pm}(k)=\frac{1}{\sqrt{2}}\sqrt{C_k\pm\sqrt{C_k^2-4 D_k^2}}
\label{SkXs}
\end{equation}
    where
\begin{eqnarray}
	& & C_k=A_0^2+B_0^2-2\cos(\theta_A+\theta_B)\cos^2 k_x
	+2(A_0\cos^2\theta_A-B_0)\cos k_y+2\cos^2\theta_A\cos^2 k_y,   \nonumber  \\
	&  & D_k^2=[(A_0+\cos k_y)(B_0-\cos k_y)-\cos^2 k_x]
	[(A_0+\cos2\theta_A\cos k_y)(B_0-\cos k_y)-\cos^2(\theta_A+\theta_B)\cos^2 k_x]
\end{eqnarray}
    where
\begin{eqnarray}
	& & A_0=h\cos\theta_A+\cos2\theta_A-\cos(\theta_A+\theta_B)>0;  \nonumber  \\
	&  & B_0=h\cos\theta_B+1-\cos(\theta_A+\theta_B)>0;
\end{eqnarray}

    We now evaluate $D_k^2$ at the $\Gamma=(0,0)$ point,
\begin{equation}
	D_k^2\big|_{k=0}=
	[(A_0+1)(B_0-1)-1]
	[(A_0+\cos2\theta_A)(B_0-1)-\cos^2(\theta_A+\theta_B)]
\end{equation}

    Note that Eqn.\ref{saddleSkX} leads to $ (A_0+1)(B_0-1)-1=0 $, therefore $ D_k^2\big|_{k=0}=0 $.
    It indicates $ \omega_{-}(k=0) $ which is the gapless Goldstone mode at $ k=0 $.

    Now we perform a long wavelength expansion around the $\Gamma$ point,
\begin{eqnarray}
	& & C_k=C_0+C_xk_x^2+C_yk_y^2+\cdots
	=C_0+2\cos(\theta_A+\theta_B)k_x^2
	+(B_0-A_0\cos^2\theta_A-2\cos^2\theta_A)k_y^2+\cdots     \nonumber  \\
	&  &D_k^2=D_x k_x^2+ D_y k_y^2+\cdots
\end{eqnarray}
    where we have introduced
\begin{eqnarray}
	& &C_0=C_k\big|_{k=0}=A_0^2+B_0^2-2\cos(\theta_A+\theta_B)
	+2(A_0\cos^2\theta_A-B_0)+2\cos^2\theta_A>0   \nonumber \\
	&  & D_x=(A_0+\cos2\theta_A)(B_0-1)-\cos^2(\theta_A+\theta_B)  \nonumber  \\
	&   &D_y=(A_0-B_0+2)[(A_0+\cos2\theta_A)(B_0-1)-\cos^2(\theta_A+\theta_B)]/2
\end{eqnarray}
     thus we can extract the Goldstone mode from $ \omega_{-}(k) $ in the long wavelength limit at the $ \Gamma=(0,0) $ point:
\begin{equation}
	\omega_{G}( \vec{k} )=\sqrt{v_{G,x}^2k_x^2+v_{G,y}^2k_y^2}
\label{goldG}
\end{equation}
     where its velocity $ v_{G,x}^2=\frac{D_x}{C_0}, v_{G,y}^2=\frac{D_y}{C_0} $ are shown in Fig.4b.

      In fact, as shown in Fig.4a, putting $ \theta_A= 0, \theta_B=\pi $ and  $ \theta_A= \theta_B=0 $, one can also push the calculations
  to the $ Z-x $ state at $ h< h_{c1} $ and the FM state at $ h > h_{c2} $, but in a different gauges than the original one used in the previous
  sections. As expected, the minimum positions of excitations may shift at different gauges. The gaps along the whole line $ \beta=\pi/4 $
  are shown in Fig.4c.

   At the lower critical field $h=h_{c_1}=\sqrt{3}-1$, $\theta_A=0$, $\theta_B=\pi$, $A_0=2+h_{c_1}$, $B_0=2-h_{c_1}$,
   we find $ v_{G,x}(h_{c1})=v_{G,y}(h_{c1})=0 $. Then we expand Eqn.\ref{SkXs} to next leading order $ k^4 $:
\begin{equation}
	\omega_{G}(\vec{k} )
	=\sqrt{\frac{(k_x^2+\sqrt{3}k_y^2)^2}{16}+O(k_x^5,k_y^5)}
	\approx\frac{k_x^2}{2m_{G1,x}}+\frac{k_y^2}{2m_{G1,y}}
\label{lowerhc1}
\end{equation}
    where we identify the two effective masses $m_{G1,x}=2, m_{G1,y}=2/\sqrt{3}$ which match
    $  m_{Z,x}, m_{Z,y} $ in Eqn.\ref{modezx} achieved from below $ h_{c1} $.
    This match indicates the transition from the $ Z-x $ state to the IC-SkX at $ h=h_{c1} $ is a second order transition with $ z=2 $.
    As shown in Fig.3(b2), in contrast to near $ h_{c2} $ to be discussed below, there is no extra roton mode near $ h_{c1} $.

    At $h=h_{c_2}=\sqrt{2}$, $\theta_A=\theta_B=0$, $A_0=B_0=h_{c_2}$, we find $ v_{G,x}(h_{c2})=v_{G,y}(h_{c2})=0 $. Then we
    expand Eqn.\ref{SkXs} to next leading order $ k^4 $:
\begin{equation}
	\omega_{G}( \vec{k} )
	=\sqrt{\frac{(k_x^2+k_y^2)^2}{4}+O(k_x^5,k_y^5)}
	\sim \frac{k_x^2}{2m_{G2,x}}+\frac{k_y^2}{2m_{G2,y}}
\label{modeSkX0}
\end{equation}
    where we identify the two effective masses $m_{G2,x}=m_{G2,y}=1$ and $ z=2 $.

    As shown in the Fig.3(b1), as $ h \rightarrow h^{-}_{c2} $, there is also roton mode   developing at $ (0, \pi ) $.
    Expanding  Eqn.\ref{SkXs} near $k=q+(0,\pi)$ leads to:
\begin{align}
	\omega_{\pm}(q)=\frac{1}{\sqrt{2}}\sqrt{ C^{\prime}_q \pm \sqrt{C^{\prime 2}_q-4 D^{\prime 2}_q } }
\end{align}
     where
\begin{align}
	&C^{\prime}_q=A_0^2+B_0^2-2\cos(\theta_A+\theta_B)\cos^2 q_x
	-2(A_0\cos^2\theta_A-B_0)\cos q_y+2\cos^2\theta_A\cos^2 q_y, \nonumber  \\
	&D^{\prime 2}_q=[(A_0-\cos q_y)(B_0+\cos q_y)-\cos^2 q_x]
	[(A_0-\cos2\theta_A\cos q_y)(B_0+\cos q_y)
	 -\cos^2(\theta_A+\theta_B)\cos^2 q_x]
\end{align}

    Now we perform a long wavelength expansion around $ (0, \pi ) $:
\begin{align}
	&C^{\prime}_q=C'_0+C'_x q_x^2+C'_y q_y^2 + \cdots   \nonumber  \\
	&D^{\prime 2}_q=D'_0+D'_x q_x^2+D'_y q_y^2+ \cdots
\end{align}
    where we have introduced:
\begin{align}
	&C'_0=A_0^2+B_0^2-2\cos(\theta_A+\theta_B)
	-2(A_0\cos^2\theta_A-B_0-\cos^2\theta_A)   \nonumber  \\
	&C'_x=2\cos(\theta_A+\theta_B)   \nonumber  \\
	&C'_y=A_0\cos^2\theta_A-B_0-2\cos^2\theta_A
\end{align}
and
\begin{align}
	&D'_0=[(A_0-1)(B_0+1)-1]
	[(A_0-\cos2\theta_A)(B_0+1)-\cos^2(\theta_A+\theta_B)]   \nonumber  \\
	&D'_x=(A_0-\cos2\theta_A)(B_0+1)-\cos^2(\theta_A+\theta_B)  \nonumber  \\
	&D'_y=(2-A_0+B_0)[(A_0-\cos2\theta_A)(B_0+1)-\cos^2(\theta_A+\theta_B)]/2
\end{align}

    Thus the roton mode in $ \omega_{-}(q) $ takes the  form:
\begin{equation}
	\omega_{R}( \vec{q} )
	=\sqrt{\Delta^2_R+v_{R,x}^2 q_x^2+v_{R,y}^2 q_y^2
	+\frac{q_x^4}{4m_{R,x}^2}
	+\frac{q_y^4}{4m_{R,y}^2}
	+\frac{q_x^2q_y^2}{2m_{R,xy}^2}},
\label{rotonR}
\end{equation}
where
\begin{align}
	\Delta_R=\sqrt{\big(C'_0-\sqrt{C_0^{\prime 2}-4D'_0}\big)/2},\quad
	v_{R,x}^2
	=
	\Big[C'_x-\frac{C'_0C'_x-2D'_x}{\sqrt{C_0^{\prime 2}-4D'_0}}\Big]\Big/2,\quad
	v_{R,y}^2
	=
	\Big[C'_y-\frac{C'_0C'_y-2D'_y}{\sqrt{C_0^{\prime 2}-4D'_0}}\Big]\Big/2.
\end{align}

    Near the upper critical field $ h \rightarrow h^{-}_{c2} $, we find $ \Delta_R =\frac{4}{7}\sqrt{10+\sqrt{2}}\big(h_{c2}-h\big)+O[(h_{c2}-h)^2] $.
    At $ h=h{c_2}=\sqrt{2}, \Delta_R=0 $, $\theta_A=\theta_B=0$, $ v_{R,x}=v_{R,y}=0 $, then the roton mode  becomes critical:
\begin{equation}
	\omega_{R}(q)
	=\sqrt{\frac{(q_x^2+q_y^2)^2}{4}+O(q_x^5,q_y^5)}
	 \sim \frac{q_x^2}{2m_{R2,x}}+\frac{q_y^2}{2m_{R2,y}}
\label{modeSkX1}
\end{equation}
   where we also identify the two effective masses $m_{R2,x}=m_{R2,y}=1$ and  $ z=2 $.

   The effective masses of both the Goldstone mode Eqn.\ref{modeSkX0} and the roton mode  Eqn.\ref{modeSkX1}
   coincide with the $ m_{F,x}, m_{F,y} $ achieved from the FM state Eqn.\ref{modefm}, or equivalently  $ h \rightarrow h^{+}_{c2}, \beta=\pi/4 $:
   $ m_{F,x}=\sqrt{\sin^42\beta-\cos^22\beta}= m_{F,y}=\sin^22\beta/m_{F,x}=1 $ at $ \beta=\pi/4 $.

\subsection{ In-commensurate SkX when $ \beta \neq \pi/4 $  }

   Eqn.\ref{icphase}  can be easily generalized to the most general IC-SkX at general $ \beta $:
\begin{eqnarray}
	&S_i=S(\sin\theta_A\cos(\phi_A-i_yk^{0}_{y}),
	      \sin\theta_A\sin(\phi_A-i_yk^{0}_{y}),
	      \cos\theta_A),\quad i\in A;   \nonumber  \\
	&S_j=S(\sin\theta_B\cos(\phi_B+j_yk^{0}_{y}),
	      \sin\theta_B\sin(\phi_B+j_yk^{0}_{y}),
	      \cos\theta_B),\quad j\in B;
\label{icphaseany}
\end{eqnarray}
   At $ \beta =\pi/4 $, $ k^{0}_{y}= \pi/2 $, Eqn.\ref{icphaseany} reduces to Eqn.\ref{icphase}.

   Because $ \phi_A + \phi_B=0 $, one can set $\phi_A=-\phi_B=\phi $
   which is nothing but the gapless Goldstone mode.
   Then Eqn.\ref{icphaseany} can be cast into the form in Eqn.M11:
\begin{eqnarray}
   S^{z} & = & A + B (-1)^{x}
       \nonumber  \\
   S^{+} & = & [S^{-}]^{\dagger}=[ C + D (-1)^{x} ] e^{i (-1)^{x} [ k^{0}_{y} y + \phi ] }
\label{icphaseform}
\end{eqnarray}
   where $ A, B $ and $ C, D $ can be expressed in terms of $ \theta_A $ and $ \theta_B $.

   The Classic ground state energy of the IC-SkX becomes
\begin{align}
	E_{\rm classic}
	=-NJS^2\big[
	(\cos^2\theta_A+\cos^2\theta_B)/2
	+(\cos(2\beta+k^{0}_{y})\sin^2\theta_A
	 +\cos(2\beta-k^{0}_{y})\sin^2\theta_B)/2\nonumber\\
	-\cos(\theta_A+\theta_B)
	+h(\cos\theta_A+\cos\theta_B)\big]
\end{align}
   which reduces to Eqn.\ref{classpi4} at $ \beta=\pi/4 $.

   The classical $\theta_A, \theta_B$ and $k^{0}_{y}$ are determined by the minimization condition:
\begin{eqnarray}
	&-\frac{1}{NJS^2}
	\frac{\partial E_{\rm classic}}{\partial\theta_A}
	=[\cos(2\beta+k^{0}_{y})-1]\sin\theta_A\cos\theta_A
	+\sin(\theta_A+\theta_B)-h\sin\theta_A
	=0   \nonumber  \\
	&-\frac{1}{NJS^2}
	\frac{\partial E_{\rm classic}}{\partial\theta_B}
	=[\cos(2\beta-k^{0}_{y})-1]\sin\theta_B\cos\theta_B
	+\sin(\theta_A+\theta_B)-h\sin\theta_B
	=0   \nonumber  \\
	&-\frac{1}{NJS^2}
	\frac{\partial E_{\rm classic}}{\partial k^{0}_{y}}
	=-[\sin(2\beta+k^{0}_{y})\sin^2\theta_A
	   -\sin(2\beta-k^{0}_{y})\sin^2\theta_B]/2
	=0
\end{eqnarray}
    which reduces to Eqn.\ref{saddleSkX} at $ \beta=\pi/4 $.
   Along the horizontal line $ h=1 $, they are shown in Fig.5a.
   At $ h< h_L, \theta_A=\theta_B, k^{0}_y=0 $, it is in the canted phase in the left of Fig.2.
   At $ h > h_R, \theta_A=\theta_B, k^{0}_y=\pi $, it is in the canted phase in the right of Fig.2.
   Obviously, there is a mirror symmetry about $ \beta=\pi/4 $ in Fig.5a.

   Following similar procedures as those at $ \beta=\pi/4 $  outlined in IV-A:
   from Eqn.\ref{icphaseany}, we make suitably chosen rotations $ \tilde{S}_i=R_y(\theta_i)R_z(\phi_i) S_i $ where
   $ \phi_i=\phi-i_y k^{0}_{y},~\theta_i=\theta_A, i \in A; \phi_j=-\phi+j_y k^{0}_{y},~\theta_j=\theta_B,j\in B $
   to align the spin quantization axis along the $Z$ axis. Then one need only introduce two HP bosons $a/b$ for the
   two sublattices A/B respectively and perform a Bogoliubov transformation to obtain the spin wave spectrum $ \omega_{\pm}(k) $.
   After very lengthy manipulations and very careful long wavelength expansion,
   we find the Goldstone mode at $ \Gamma=(0,0) $ in Eqn.\ref{goldG} at $ \beta=\pi/4 $ is replaced by:
\begin{equation}
	\omega_{G}( \vec{k} )=\sqrt{v_{G,x}^2k_x^2+v_{G,y}^2k_y^2}-c_G k_y
\label{goldGca}
\end{equation}
    where $ c_{G} ( \beta, H )= -c_G( \pi/2-\beta, H ) $, so $ c_G > 0 $ when $ \beta < \pi/4 $,
    $ c_G < 0 $ when $ \beta > \pi/4 $ and $ c_G = 0 $ when $ \beta = \pi/4 $ recovering Eqn.\ref{goldG}.
    How the three velocities $ v_{G,x}, v_{G,y} $ and $ c_G $ changes from $ h_{c1} $ to $ h_{c2} $ at a fixed $ \beta= \pi/5 < \pi/4 $ is shown in
    Fig.5b.

    Similarly, the roton mode  Eqn.\ref{rotonR} at $ \beta=\pi/4 $ developed as $ h \rightarrow h^{-}_{c2} $ near $ ( 0, \pi) $ is replaced by:
\begin{equation}
	\omega_{R}( \vec{q} )
	=\sqrt{\Delta^2_R+v_{R,x}^2 q_x^2+v_{R,y}^2 q_y^2}-c_R q_y
\label{rotonRca}
\end{equation}
    where $ c_{R} ( \beta, H )= -c_R( \pi/2-\beta, H ) $, so $ c_R > 0 $ when $ \beta < \pi/4 $,
    $ c_R < 0 $ when $ \beta > \pi/4 $ and $ c_R = 0 $ when $ \beta = \pi/4 $ recovering Eqn.M11.

    Compared with Eqn.M6 and M7, we find the Goldstone mode and the Roton mode take similar forms as those in the canted phase.
    At a fixed $ h $ in Fig.2 ( for fixed $ h=1 $, see Fig.5a ), we find that as $ h \rightarrow h^{+}_{L} $
    ( or $ h \rightarrow h^{-}_{R} $ ) , $ v_{G,y}- c_G \rightarrow 0 $ ( or $ v_{G,y}+ c_G \rightarrow 0 $ ),
    it is a bosonic Lifshitz transition with the anisotropic dynamic exponent $ z_x=1, z_y=3 $.
    This picture is completely consistent as that achieved from the canted phase to the IC-SkX.
    These facts suggest some sort of duality between the cant phase and the IC-SkX phase on the two side of $ h_L $ in Fig.1.

    Taking $ h \rightarrow h^{+}_{c1} $, $ v_{G,x}=v_{G,y}=0 $ and $ c_G=0 $ in Eqn.\ref{goldGc}, expanding it to the order $ k^{4} $, we find it
    matches Eqn.M2 reached from $ Z-x $ state below $ h_{c1} $.

    Taking $ h \rightarrow h^{-}_{c2} $, $ v_{G,x}=v_{G,y}=0 $ and $ c_G=0 $ in Eqn.\ref{goldGc} and
    $ \Delta_R=0, v_{R,x}=v_{R,y}=0 $ and $ c_R=0 $ in Eqn.\ref{rotonRc}, expanding both equations to order $ k^{4} $, we find both
    matches $ m_{F,x} $ and $ m_{F,y} $ reached from FM state above $ h_{c2} $.

\end{widetext}


\begin{thebibliography}{99}
\bibitem{comp}
Note that in this paper, we only studied the RFHM+H on a square lattice, it is important to extend it to
a honeycomb lattice \cite{tqpt,S:un}
with generic gauge parameters $ (\alpha,\beta,\gamma) $ put on the three bonds.
Then the IC-SkX phase may be turned into a spin liquid phase which breaks no symmetries of the Hamiltonian, but with some
topological orders.
\bibitem{skxdensity} Obviously, the Skyrmion density $ Q_{ijk} $ is not gauge invariant. In the $ U(1)_{soc} $ basis, it becomes
 a constant $ Q_{ijk}= -\sin (\theta_A- \theta_B) \sin \theta_B $ independent of the lattice site and the angle $ \phi $.
\bibitem{rafm2} It is quite non-trvivial and difficult to get a higher spin Rotated Anti-ferromagnet (RAFH) studied in \cite{rafm}.
So it is quite difficult to raise the critical temperatures of RAFH.
On the cold atom experimental side, the heating issue with fermions is also more serious than that of spinor bosons.
The two factors combine to make the observation of low temperaure phases of RAFH more difficult than those of RFH studied in this paper.
\end{thebibliography}
\end{document}